\documentclass[prd,twocolumn,nopacs,floatfix,amsmath,nofootinbib,amssymb,floatfix]{revtex4}
\usepackage{graphicx,color,dcolumn,booktabs,bm}
\usepackage{longtable,lscape}
\usepackage{txfonts}
\usepackage{overpic}
\usepackage{diagbox}
\usepackage{multirow}
\usepackage{gensymb}
\usepackage{booktabs}
\usepackage{amssymb}
\usepackage{indentfirst}
\usepackage{feynmf}   
\usepackage{slashed}  
\usepackage{cases}
\usepackage{color}
\usepackage{epstopdf}
\usepackage{graphicx,color,dcolumn,booktabs,bm}
\usepackage[colorlinks, citecolor=blue,anchorcolor=red,menucolor=red, linkcolor=red,filecolor=red,runcolor=red,urlcolor=blue,frenchlinks=red]{hyperref}

\graphicspath{{Figures/}} %

\begin{document}

\title{Extending the Constituent Gluon Model to Heavy-Flavour Hybrids: A Unified Study of $c\bar{c}g$ Mesons}

\author{Qi Huang$^{1,6}$}
\author{Zhe-Tao Miu$^1$}
\author{Rui Chen$^{2,5,6}$}\email{chenrui@hunnu.edu.cn}
\author{Xiao-Huang Hu$^3$}
\author{Yue Tan$^4$}
\affiliation{
$^1$Department of Physics and Technology, Nanjing Normal University, Nanjing 210023, China\\
$^2$Key Laboratory of Low-Dimensional Quantum Structures and Quantum Control of Ministry of Education, Department of Physics and Synergetic Innovation Center for Quantum Effects and Applications, Hunan Normal University, Changsha 410081, China\\
$^3$Department of Physics, Changzhou Vocational Institute of Engineering, Changzhou 213164, China\\
$^4$Department of Physics, Yancheng Institute of Technology, Yancheng 224000, China\\
$^5$Hunan Research Center of the Basic Discipline for Quantum Effects and Quantum Technologies, Hunan Normal University, Changsha 410081, China\\
$^6$Lanzhou Center for Theoretical Physics, Key Laboratory of Theoretical Physics of Gansu Province, Lanzhou University, Lanzhou 730000, China
}

\begin{abstract}
We investigate the mass spectra and two-body strong decay properties of ground charmonium hybrids within the framework of a constituent gluon model. Based on the assumption that non-perturbative QCD endows the gluon with an effective mass, we extend the chiral quark model by introducing a single new parameter, the constituent gluon mass $m_g=450$~MeV, which is fixed from previous studies of light hybrids, while other parameters are taken directly from successful descriptions of ordinary meson spectra. We systematically compute the spectra for various quantum numbers and find good agreement with results from lattice QCD, potential models, and other approaches. The corresponding decay widths are also reasonable. For experimental searches, we recommend focusing on the exotic $1^{-+}$ and $2^{+-}$ states, which decay prominently into $D\bar{D}_1$ and $D\bar{D}_2^*$ channels, respectively. Among ordinary quantum numbers, the $0^{-+}$, $2^{-+}$, and $1^{+-}$ states with significant decays into orbitally excited charm mesons are also suggested. Our results provide a unified and consistent description of charmonium hybrids and offer clear guidance for future experimental identification.
\end{abstract}

\maketitle

\section{Introduction}
As the fundamental theory describing interactions involving quarks and gluons, Quantum Chromodynamics (QCD) predicts not only conventional mesons ($q\bar{q}$) and baryons ($qqq$) but also a rich spectrum of exotic hadrons, including multiquark states, glueballs, and hybrids. Among these, the hybrid state is particularly notable, as it contains not only quark constituents but also gluonic fields, rendering studies of such systems essential for understanding the dynamical role of gluonic degrees of freedom.

Despite the successful description of quark degrees of freedom in current phenomenological models such as the potential model, the understanding of the gluon---theoretically a massless gauge boson in QCD---remains under investigation, constituting the central problem in decoding hybrid states. For this issue, Refs.~\cite{Binosi:2012sj,Binosi:2022djx,Ding:2022ows} propose that gluons can acquire an effective mass via the Schwinger mechanism, where non-perturbative effects bring a mass-like term with $m_g\approx450$ MeV into the gluon propagator. Based on this conclusion, a straightforward idea emerges: explicit gluons may be treated as constituents just like valence quarks, implying that as long as an extended potential model that contains a quark-gluon interaction is available, the hybrid system can then be treated as a calculable few-body system.

To verify this proposal, the simplest hybrid---the hybrid meson composed of a quark-antiquark pair and an explicit gluon---is the optimal choice, as the constituent-gluon model transforms such a system into a three-body problem. In a recent work, we applied this treatment to the $J^{PC}=1^{-+}$ light hybrid meson $q\bar{q}g$ and found that using exactly the same model parameters determined from ordinary meson spectroscopy yields, the obtained behaviors of such hybrid are consistent with previous theoretical studies, particularly for the ground states \cite{Bernard:2003jd,Dudek:2013yja,Tan:2024grd,Meyer:2015eta,Shastry:2022mhk,Zhang:2025xee,Ma:2025cew}. These results demonstrate that once the constituent gluon mass is properly included, the same model describing ordinary mesons can naturally reproduce the light hybrid spectrum, a manifestation of the expected universality of QCD.

However, to further examine the effectiveness of our constituent gluon model, the study of hybrids should not be limited to the light sector with just a single quantum number. Thus, it is necessary to extend our model to the charm sector with the same constituent gluon mass $m_g=450$ MeV, so that the obtained results can be used to further test the universality of the model parameters and offer an important complementary reference to first-principles methods such as lattice QCD.

In addition, previous studies of charmonium hybrids can be grouped into several broad classes~\cite{Qian:2026inl,Cheung:2016bym,Farina:2020slb,HadronSpectrum:2012gic,Wang:2025clb,Akbar:2024jda,Vairo:2017dmh,Lebed:2017xih,TarrusCastella:2017lex,TarrusCastella:2015mbn,Berwein:2015vca,Braaten:2014qka,Kleiv:2014kua,Sultan:2014oua,Barnes:1982tx,Chanowitz:1982qj,Barnes:1995hc,Kalashnikova:2016bta,Agaev:2025llz,Chen:2013zia,Oncala:2017hop,Iddir:2007dq,Iddir:2006sh,Wang:2025dur,Buisseret:2006wc,Barnes:1981ac}. Lattice QCD calculations~\cite{Cheung:2016bym,HadronSpectrum:2012gic} provide first-principles spectra. QCD sum rules~\cite{Kleiv:2014kua,Agaev:2025llz,Chen:2013zia} mainly aim at mass determinations. The bag model~\cite{Barnes:1982tx,Chanowitz:1982qj} and flux-tube model~\cite{Barnes:1995hc} give mass spectra and, in the flux-tube case, also decay amplitudes. Constituent-gluon models~\cite{Qian:2026inl,Farina:2020slb,Iddir:2007dq,Iddir:2006sh} treat the gluon as an effective degree of freedom and often compute both masses and two-body decay widths. Born–Oppenheimer and  related effective-field-theory approaches~\cite{Vairo:2017dmh,TarrusCastella:2017lex,TarrusCastella:2015mbn,Berwein:2015vca,Braaten:2014qka,Oncala:2017hop} describe hybrids in terms of static potentials and yield mass splittings. Extended potential models~\cite{Akbar:2024jda,Sultan:2014oua} incorporate gluonic excitations into the quark-antiquark  potential and determine hybrid masses by solving the  radial Schr\"odinger equation using power-series or numerical shooting methods. The QCD string model~\cite{Kalashnikova:2016bta} provides another description of confinement and has been used to calculate the masses of heavy-quark hybrids. The relation between the constituent gluon and excited-string pictures has also been explored~\cite{Buisseret:2006wc}, using the auxiliary field method to derive approximate analytical mass spectra in both three-body and two-body descriptions. Hybrid decays have also been studied within constituent gluon models~\cite{Qian:2026inl,Farina:2020slb,Iddir:2007dq,Iddir:2006sh}, QCD sum rules~\cite{Agaev:2025llz}, and the pNRQCD framework~\cite{Oncala:2017hop}.

In this context, the present work provides a systematic and simultaneous calculation of both the ground spectrum and the complete decay patterns of charmonium hybrids, thereby offering a new independent calculation for comparison and helping to distinguish and reconcile the differences among various models regarding the properties of hybrids. Furthermore, as a Fock component of charmonium under the unquenched quark model, mixing between charmonia and $c\bar{c}g$ hybrids must exist, leading to mass shifts. Thus, an early investigation of the behavior of charmonium hybrids may be helpful for further studies of the unquenched effects of charmonia.

On the experimental side, the BESIII Collaboration has searched for $1^{-+}$ charmonium-like hybrids in the energy range 4.258--4.681 GeV \cite{BESIII:2025bez,Wang:2025dur}, reflecting experimental interest in this subject in the charm sector. By predicting specific decay channels with significant branching fractions, this work can provide direct and testable search targets for experiments such as PANDA, BESIII, and Belle-II, thereby offering clear guidance for the eventual experimental identification of charmonium hybrids.

The paper is organized as follows. In Sec.~\ref{sec2}, we introduce the theoretical framework. Then, the numerical results and discussion are presented in Sec.~\ref{sec3}. Finally, the paper concludes with a summary.

\section{Theoretical framework}
\label{sec2}

\subsection{Potential model}

When discussing interactions between quarks, the QCD-inspired chiral quark model remains one of the most commonly used approaches, which has been successfully applied to hadron spectra~\cite{Vijande:2004he,Segovia:2008zza,Segovia:2008zz,Ortega:2016hde}, hadron-hadron interactions~\cite{Fernandez:1993hx,Valcarce:1994nr,Ortega:2016mms,Ortega:2016pgg}, and multiquark structures~\cite{Vijande:2006jf,Yang:2020atz,Huang:2023jec}. In the present work, since we treat the gluon as a constituent, we must extend the chiral quark model to incorporate the interaction between the valence quark and the gluon, yielding the Hamiltonian for the hybrid meson as
\begin{eqnarray}
\label{eq1}
  \hat{H} &=& \frac{\hat{p}^2_{q,\bar{q}}}{2\mu_{q,\bar{q}}} + \frac{\hat{p}^2_{q\bar{q},g}}{2\mu_{q\bar{q},g}} + \hat{V}_{q\bar{q}} + \hat{V}_{qg} + \hat{V}_{\bar{q}g},
\end{eqnarray}
where $\hat{p}^2_{q,\bar{q}}$ and $\mu_{q,\bar{q}}$ are the relative momentum and reduced mass between $q$ and $\bar{q}$, while $\hat{p}^2_{q\bar{q},g}$ and $\mu_{q\bar{q},g}$ are the relative momentum and reduced mass between the $q\bar{q}$ cluster and the constituent gluon.

Within the original chiral quark model, the interaction between a quark $q$ and an antiquark $\bar{q}$ takes the form
\begin{equation}
\begin{split}
  V_{q\bar{q}}(r) = V^{\text{CON}}_{q\bar{q}}(r) + V^{\text{OGE}}_{q\bar{q}}(r) + V^{\text{GBE}}_{q\bar{q}}(r),
\end{split}
\end{equation}
where $V^{CON}_{q\bar{q}}$, $V^{OGE}_{q\bar{q}}$, and $V^{GBE}_{q\bar{q}}$ represent the color confinement, one-gluon-exchange, and Goldstone-boson-exchange potentials, respectively. These three terms capture the most essential features of low-energy QCD: color confinement, asymptotic freedom, and spontaneous chiral symmetry breaking~\cite{Vijande:2004he}. For color confinement, we include both the central screened form\footnote{As shown in our previous work~\cite{Ma:2025cew}, as long as the meson spectra are well fitted, the changes to the spectra of ground hybrids remain small.} and the Thomas-precession terms as
\begin{eqnarray}
   V_{q\bar{q}}^{\mathrm{CON},C}(\boldsymbol{r}_{ij}) &=&
   \left( -\boldsymbol{\lambda}_i^c \cdot \boldsymbol{\lambda}_j^{c*} \right)
   \left[ -a_c \left( 1 - e^{-\mu_c r_{ij}} \right) + \Delta \right],\\
   V_{q\bar{q}}^{\mathrm{CON},SO}(\boldsymbol{r}_{ij}) &=&
   - \left( -\boldsymbol{\lambda}_i^c \cdot \boldsymbol{\lambda}_j^{c*} \right)
   \frac{a_c \mu_c e^{-\mu_c r_{ij}}}{4 m_i^2 m_j^2 r_{ij}}
   \Big[
   \Big( (m_i^2 + m_j^2)\nonumber\\
   &&\times(1 - 2a_s) + 4 m_i m_j (1 - a_s) \Big) (\boldsymbol{S}_+ \cdot \boldsymbol{L})\nonumber\\
   &&+ (m_j^2 - m_i^2)(1 - 2a_s) (\boldsymbol{S}_- \cdot \boldsymbol{L})\Big],
\end{eqnarray}
where $a_c$, $\mu_c$, $\Delta$, and $a_s$ are model parameters, and $\lambda^c$ represents the $\text{SU}(3)$ color Gell-Mann matrices. While for the one gluon exchange potential contains Coulomb, color-magnetic, spin-orbit, and tensor interactions as
\begin{eqnarray}
   V_{q\bar{q}}^{\mathrm{OGE},C}(\boldsymbol{r}_{ij}) &=&
   \left( -\boldsymbol{\lambda}_i^c \cdot \boldsymbol{\lambda}_j^{c*} \right)
   \frac{\alpha_s}{4}
   \Big[
   \frac{1}{r_{ij}} - \frac{1}{6m_i m_j}
   \frac{e^{-r_{ij}/r_0(\mu)}}{r_{ij} r_0^2(\mu)}\nonumber\\
   &&\times\left( \boldsymbol{\sigma}_i \cdot \boldsymbol{\sigma}_j \right)\Big],\\
   V_{q\bar{q}}^{\mathrm{OGE},SO}(\boldsymbol{r}_{ij}) &=&
   \left( -\boldsymbol{\lambda}_i^c \cdot \boldsymbol{\lambda}_j^{c*} \right)
   \frac{-\alpha_s}{16m_i^2 m_j^2}
   \Big[
    \frac{1}{r_{ij}^3} - \frac{e^{-r_{ij}/r_g(\mu)}}{r_{ij}^3} \\
   &&\times \left( 1 + \frac{r_{ij}}{r_g(\mu)} \right)
   \Big]
   \Big[
   \left( (m_i + m_j)^2 + 2m_i m_j \right) \nonumber\\
   &&\times(\boldsymbol{S}_+ \cdot \boldsymbol{L})
   + (m_j^2 - m_i^2) (\boldsymbol{S}_- \cdot \boldsymbol{L})
   \Big], \\
   V_{q\bar{q}}^{\mathrm{OGE},T}(\boldsymbol{r}_{ij}) &=&
   \left( -\boldsymbol{\lambda}_i^c \cdot \boldsymbol{\lambda}_j^{c*} \right)
   \frac{-\alpha_s}{16m_i m_j}
   \Big[
   \left( \frac{1}{r_{ij}^3} - \frac{e^{-r_{ij}/r_g(\mu)}}{r_{ij}} \right)\nonumber\\
   &&\times
   \Big(
   \frac{1}{r_{ij}^2} + \frac{1}{3r_g^2(\mu)} + \frac{1}{r_{ij}r_g(\mu)}
   \Big)\Big] \boldsymbol{S}_{ij}.
\end{eqnarray}
Here, $\mu$ is the reduced mass of the interacting quark pair, $\sigma$ represents the Pauli spin matrices, $\boldsymbol{S}_{ij}$ is the tensor operator, $r_0(\mu)=\hat{r}_0/\mu$ and $r_g(\mu)=\hat{r}_g/\mu$ define the scale parameters dependent on the reduced mass, and $\alpha_s$ denotes the effective scale-dependent running strong coupling constant parametrized by $\alpha_0$, $\mu_0$ and $\Lambda_0$ as~\cite{Vijande:2004he}
\begin{equation}
  \alpha_s = \frac{\alpha_0}{\log\left( \frac{\mu^2 + \mu_0^2}{\Lambda_0^2} \right)}.\label{eq:runalphas}
\end{equation}

Apart from the confinement and one-gluon-exchange terms, and owing to chiral symmetry breaking, additional Goldstone-boson-exchange potentials also emerge in the interaction between light quarks ($u$, $d$, and $s$) as
\begin{eqnarray}
V^{\mathrm{GBE}}_{q\bar{q}}(\boldsymbol{r}_{ij})&=&V_{\pi}(\boldsymbol{r}_{ij})\sum_{a=1}^{3} \lambda_i^a \lambda_j^{a*}+V_{K}(\boldsymbol{r}_{ij})\sum_{a=4}^{7} \lambda_i^a \lambda_j^{a*} \nonumber\\
&&+V_{\eta}(\boldsymbol{r}_{ij})\Big[ \cos\theta_P \left( \lambda_i^8 \lambda_j^{8*} \right)- \sin\theta_P \left( \lambda_i^0 \lambda_j^{0*} \right) \Big] \nonumber\\
&&+V_{\sigma}(\boldsymbol{r}_{ij}),
\end{eqnarray}   
where the detailed expressions read
\begin{align}
    V_{\chi=\pi,K,\eta}({{\boldsymbol{r}}_{ij}}) = & \frac{g^2_{ch}}{4\pi}\frac{m^2_\chi}{12m_im_j} \frac{\Lambda^2_\chi }{\Lambda^2_\chi-m^2_\chi} m_\chi\left\{\Bigg[ Y(m_{\chi}r_{ij}) \right.\nonumber \\
    &\left. -\frac{\Lambda^3_\chi}{m^3_\chi} Y(\Lambda_{\chi}r_{ij}) \right] (\boldsymbol{\sigma}_i \cdot \boldsymbol{\sigma}_j) + \Bigg[H\left(m_\chi r_{i j}\right)\nonumber\\
    &\left.\left.-\frac{\Lambda_\chi^3}{m_\chi^3} H\left(\Lambda_\chi r_{i j}\right)\right]\boldsymbol{S}_{i j} \right\} ,\\
    V_{\sigma}({{\boldsymbol{r}}_{ij}}) = & -\frac{g_{c h}^2}{4 \pi}\frac{\Lambda_\sigma^2}{\Lambda_\sigma^2-m_\sigma^2} m_\sigma\left\{\Bigg[Y\left(m_\sigma r_{i j}\right)\right. \nonumber\\
    &\left.-\frac{\Lambda_\sigma}{m_\sigma} Y\left(\Lambda_\sigma r_{i j}\right)\right] + \frac{m_\sigma}{2 m_i m_j}\Bigg[G\left(m_\sigma r_{i j}\right)\nonumber\\
    &\left.-\frac{\Lambda_\sigma^3}{m_\sigma^3} G\left(\Lambda_\sigma r_{i j}\right)\Bigg]\left(\boldsymbol{S}_+\cdot\boldsymbol{L}\right)\right\}.
\end{align}
Here, $\Lambda_{\chi,\sigma}$ are the cutoff parameters of the corresponding Goldstone boson exchanges, $Y(x)=e^{-x}/x$ denotes the Yukawa function, $H(x)=\left(1+\frac{3}{x}+\frac{3}{x^2}\right)Y(x)$, $G(x)=\left(1+\frac{1}{x}\right)\frac{Y(x)}{x}$, and $\lambda^a$ represents the $SU(3)$ Gell-Mann matrices. Furthermore, $m_\sigma$ is determined via the relation $m_\sigma ^2 = m_\pi ^2 + 4m_{u,d}^2,$ and $g_{ch}^2$ is the chiral field coupling constant, which is fixed from the nucleon-nucleon-pion $(NN\pi)$ coupling constant as~\cite{Vijande:2004he}
\begin{equation}
\begin{split}
   \frac{g_{ch}^2}{4\pi} = \frac{9}{25} \frac{g_{\pi NN}^2}{4\pi} \frac{m_{u,d}^2}{m_N^2}.
\end{split}
\end{equation}
   
The remaining interaction is between the (anti-)quark and the gluon, containing only the confinement and one-gluon-exchange terms:
\begin{equation}
  V_{qg}(\boldsymbol{r}_{ij}) = V^{\text{CON}}_{qg}(\boldsymbol{r}_{ij}) + V^{\text{OGE}}_{qg}(\boldsymbol{r}_{ij}).
\end{equation}
Here, the confinement term shares the same screened form as $V_{q\bar{q}}^{\mathrm{CON}}$, but with the color operator replaced by
\begin{equation}
  V^{\text{CON}}_{qg}(r) = V^{\text{CON}}_{q\bar{q}}(r) \quad \text{with} \quad \boldsymbol{\lambda}_c \cdot \boldsymbol{\lambda}_c^* \to i\boldsymbol{\lambda}^d \cdot \boldsymbol{f}^d.
\end{equation}
Here, $\boldsymbol{f}_c=(f_c)_{ab}=f_{cab}$ are the antisymmetric structure constants of $SU(3)$, while the one-gluon-exchange potential between a quark and a gluon is directly derived from the QCD Lagrangian. Its explicit form is obtained by applying a non-relativistic reduction combined with a Fourier transformation to the $t$-channel scattering amplitude of the quark-gluon interaction~\cite{Ma:2025cew,Mathieu:2007fpv,Varshalovich:1988ifq} and reads
\begin{eqnarray}
    V_{qg}(\boldsymbol{r}) &=& \frac{\alpha_s}{2}  \boldsymbol{\lambda}_c \cdot \boldsymbol{f}_c \left[\frac{1}{r}-\left(\frac{2\pi}{3m_g^2}+\frac{\pi}{2m_q^2}\right) \frac{\mu e^{-\mu r / r_0}}{4\pi r_0 r} \right.\nonumber\\
    &&+ \frac{1}{2m_g^2r^3}\left(\boldsymbol{S}_g \cdot (\boldsymbol{L}_g+ \boldsymbol{S}_g)-3\frac{(\boldsymbol{S}_g \cdot \boldsymbol{r})(\boldsymbol{S}_g \cdot \boldsymbol{r})}{r^2}\right) \nonumber\\
	&&\left.-\frac{\boldsymbol{S}_q \cdot \boldsymbol{L}_q}{2m_q^2r^3}-\frac{\boldsymbol{S}_g \cdot \boldsymbol{L}_q-\boldsymbol{S}_q \cdot \boldsymbol{L}_g}{m_gm_qr^3}\right.\nonumber\\
    &&\left.-\frac{8\pi \boldsymbol{S}_g \cdot \boldsymbol{S}_q}{3m_gm_q} \frac{\mu e^{-\mu r / r_0}}{4\pi r_0 r}\right].
\end{eqnarray}

\subsection{Wave function}

The total wave function of a hadron consists of four parts: spatial, spin, flavor, and color, respectively. Thus, it can be expressed as
\begin{equation}
   \psi_{q\bar{q}g}^J=\psi_{q\bar{q}g}^L \otimes \psi_{q\bar{q}g}^S \otimes \psi_{q\bar{q}g}^f \otimes \psi_{q\bar{q}g}^c.
\end{equation}
Here, $\psi_{q\bar{q}g}^f$ denotes the flavor wave function, which is nearly the same as that of the corresponding meson but with an additional gluon $g$. $\psi_{q\bar{q}g}^c$ is the color part, which must form a color-singlet state to satisfy color confinement and is built from the $SU(3)_c$ representations of the constituents as~\cite{Mathieu:2007fpv}
\begin{equation}
|\psi^{c}_{q\bar{q}g}\rangle  = \frac{1}{4}(\lambda^a)_{ij} |q^i\rangle\otimes |\bar{q}^j\rangle \otimes |g^a\rangle,.  
\end{equation}
The spin and spatial parts are usually coupled, and for hybrids, the mainstream coupling scheme is~\cite{Ma:2025cew}
\begin{equation}
   \psi^{J,g}_{q\bar{q}g} = \left[ \left[ \psi^{\frac{1}{2}}_q \psi^{\frac{1}{2}}_{\bar{q}} \right]^{S_{q\bar{q}}} \left[ \left[ \psi^{L_{q\bar{q},g}}_{q\bar{q},g} \psi^1_g \right]^{J_g} \psi^{L_{q\bar{q}}}_{q\bar{q}} \right]^{L_g} \right]^J,\label{eq:JJ}
\end{equation}

In other words, it resembles the "light degree of freedom" inherent in heavy quark symmetry, where the gluon field first couples the two relative angular momenta $L_{q\bar{q},g}$ and $L_{q\bar{q}}$ to form an excited constituent gluon, which subsequently couples to $S_{q\bar{q}}$ to yield the total angular momentum $J$. However, this coupling scheme is somewhat cumbersome for practical calculations, since it mixes the spin and orbital wave functions of the gluon and the $q\bar{q}$ cluster. To avoid this issue, the $L$-$S$ coupling scheme is commonly adopted, whose wave function is constructed as
\begin{equation}
   \psi^{J,LS}_{q\bar{q}g} = \left[ \left[ \left[ \psi^{\frac{1}{2}}_q \psi^{\frac{1}{2}}_{\bar{q}} \right]^{S_{q\bar{q}}} \psi^1_g \right]^S \left[ \psi^{L_{q\bar{q}}}_{q\bar{q}} \psi^{L_{q\bar{q},g}}_{q\bar{q},g} \right]^L \right]^J.
\end{equation}
Then, based on angular momentum theory, we establish a relationship between these two schemes, expressed using $6j$- and $9j$-symbols as
\begin{equation}
\begin{split}
   \psi^{J,g}_{q\bar{q}g} = &\sum_{J_{q\bar{q}},S,L} \sqrt{(2J_{q\bar{q}}+1)(2L_g+1)} (-1)^{S_{q\bar{q}}+L_{q\bar{q}}+J_g+J} \\
   &\times \begin{Bmatrix} S_{q\bar{q}} & L_{q\bar{q}} & J_{q\bar{q}} \\ J_g & J & L_g \end{Bmatrix} \sqrt{(2S+1)(2L+1)} \\
   &\times \sqrt{(2J_{q\bar{q}}+1)(2J_g+1)} \begin{Bmatrix} S_{q\bar{q}} & L_{q\bar{q}} & J_{q\bar{q}} \\ 1 & L_{q\bar{q},g} & J_g \\ S & L & J \end{Bmatrix} \\
   &\times \psi^{J,LS}_{q\bar{q}g}.
\end{split}
\end{equation}

Finally, to determine the spatial wave functions, we solve the eigenvalue problem using the Rayleigh-Ritz variational method, in which the trial spatial wave function is expanded in a set of basis functions. In this work, we again adopt the widely used Gaussian expansion method (GEM)~\cite{Hiyama:2003cu} to represent each relative-motion coordinate in the system, a method that has proven to be accurate and universal for few-body calculations~\cite{Hu:2020zwc,Yang:2020fou,Tan:2020cpu}. The core idea of the method is to expand the radial part of the orbital wave function in terms of a series of Gaussian functions as
\begin{equation}
   \psi_{q\bar{q}g}^{L} = \sum_{n=1}^{n_{\max}} c_n N_{nl} e^{-\nu_n r^2} \mathcal{Y}_l(\boldsymbol{r}),
\end{equation}
in which $Y_l(\boldsymbol{r})$ is the solid spherical harmonic function and $N_{nl}$ is the normalization constant given by
\begin{equation}
   N_{nl} = \left( \frac{2^{l+2} (2\nu_n)^{l+3/2}}{\sqrt{\pi} (2l+1)!!} \right)^{1/2}.
\end{equation}
The coefficients $c_{n}$ are the variational parameters, which are determined by the dynamics of the system. The Gaussian size parameters are chosen according to the following geometric progression
\begin{equation}
   \nu_n = \frac{1}{r_n^2}, \quad r_n = r_{\min} \left( \frac{r_{\max}}{r_{\min}} \right)^{\frac{n-1}{n_{\max}-1}},
\end{equation}
where $r_{\min}$ and $r_{\max}$ are the minimum and maximum length scales for the Gaussian basis functions, while $n_{\max}$ is the number of basis functions chosen to ensure the convergence of our results. Based on our convergence tests, the spectra of ground charmonium hybrid mesons are sufficiently converged with $r_{\min}=0.1$ fm, $r_{\max}=2$ fm, and $n_{\max}=8$.

\subsection{Two body decay mechanism}

\begin{figure}[htb]
\centering
\includegraphics[width=0.45\textwidth]{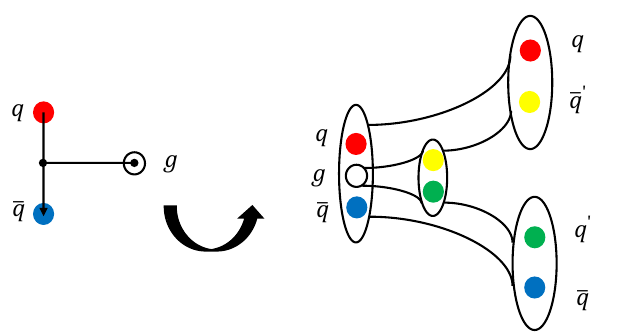} 
\caption{Schematic diagram of the hybrid meson decay mechanism, where the gluon in the hybrid (left) transitions into a $q^\prime\bar{q}^\prime$ pair, then rearrangement between this pair and the original $q^\prime\bar{q}$ happens to form the final two mesons (right).}
\label{fig:decay_diagram}
\end{figure}

For decays of the $q\bar{q}g$ hybrid meson into two mesons, the leading order transition mechanism is that the constituent gluon $g$ serves as the source for the creation of the internal $q\bar{q}$ pair, then it is followed by the rearrangement of all quarks and antiquarks into the final two mesons, as presented in Fig.~\ref{fig:decay_diagram}.
The Hamiltonian for this transition mechanism can be expressed as
\begin{equation}
   \hat{H}_I = i \sqrt{4\pi\alpha_s} \frac{(\lambda^a)_{bc}}{2} \int d^3\boldsymbol{x}\, \bar{q}^c(\boldsymbol{x}) \gamma^\mu q^b(\boldsymbol{x}) A^a_\mu(\boldsymbol{x}),
\end{equation}
in which $q^b,\bar{q}^c,$ and $A^a_\mu$ are the quark, anti-quark, and gluon fields, respectively. Through field expansions and a non-relativistic reduction~\cite{Varshalovich:1988ifq,Ding:2006ya}, we obtain the leading-order transition operator as
\begin{equation}
\begin{split}
   \hat{T} =& 3i\sqrt{\pi\alpha_s}(\lambda^a)_{bc}
   \sum_{s,s',m}
   \int \frac{d^3\boldsymbol{p}_1 d^3\boldsymbol{p}_2 d^3\boldsymbol{k}}{\sqrt{2m_g}(2\pi)^6}
   \delta(\boldsymbol{p}_1 + \boldsymbol{p}_2 - \boldsymbol{k})\\
   &\times C^{0,0}_{1,m;1,-m} C^{1/2,s}_{1,-m;1/2,s'}
   d^{c\dagger}_{s'}(\boldsymbol{p}_1) b^{b\dagger}_{s}(\boldsymbol{p}_2) a^{a}_{m}(\boldsymbol{k}),
\end{split}
\end{equation}
where $\alpha$ is the strong coupling constant, $(\lambda^a)_{bc}$ are the color matrix elements,  $a^{a}_{m}$ denotes the annihilation operator of the gluon, while $b^{b\dagger}_{s}$ and $d^{c\dagger}_{s'}$ represent the creation operators of the quark and antiquark, respectively.

Thus, the matrix element for the transition $A\to B+C$ can be calculated as
\begin{equation}
   \langle BC | T | A \rangle = \delta^3(\boldsymbol{P}_A - \boldsymbol{P}_B - \boldsymbol{P}_C) \mathcal{M}^{M_{J_A}, M_{J_B}, M_{J_C}}
\end{equation}
in which $\boldsymbol{P}_B$ and $\boldsymbol{P}_C$ denote the three-momenta of the final-state mesons $B$ and $C$ evaluated in the rest frame of the initial hybrid $A$, and $\mathcal{M}^{M_{J_A}, M_{J_B}, M_{J_C}}$ represents the helicity amplitude of the decay process that written in detail as
\begin{equation}
\begin{split}
   &\mathcal{M}^{M_{J_A}, M_{J_B}, M_{J_C}} \\
   &\quad=\sum_{\substack{M_{L_{q\bar{q},g}},M_{S_g},\\M_{L_{q\bar{q}}},M_{S_{q\bar{q}}},\\  M_{L_B},M_{S_B},\\M_{L_C},M_{S_C}}}
   \mathcal{CS}
   \left\langle L_{q\bar{q},g},M_{L_{q\bar{q},g}};1,M_{S_g}\middle|J_g,M_{L_{q\bar{q},g}}+M_{S_g}\right\rangle\\
   &\qquad\times \left\langle L_{q\bar{q}},M_{L_{q\bar{q}}};J_g,M_{J_g}\middle|L_g,M_{L_g}\right\rangle\\
   &\qquad\times \left\langle L_g,M_{L_g};S_{q\bar{q}},M_{S_{q\bar{q}}}\middle|J_A,M_{J_A}\right\rangle \\
   &\qquad\times\left\langle L_B,M_{L_B};S_B,M_{S_B}\middle|J_B,M_{J_B}\right\rangle\\
   &\qquad\times\left\langle L_C,M_{L_C};S_C,M_{S_C}\middle|J_C,M_{J_C}\right\rangle\\
   &\qquad \times \left[(-1)^{1+S_{q\bar{q}}+S_B+S_C} \left\langle\phi_B^{32}\phi_C^{14} \middle|\phi_A^{12}\phi_0^{34}\right\rangle\mathcal{I}^{(13)}\right.\\
   &\quad\quad\left.+\left\langle\phi_B^{14}\phi_C^{32}\middle|\phi_A^{12}\phi_0^{34}\right\rangle\mathcal{I}^{(24)}\right].
\end{split}
\end{equation}
Here, $\mathcal{C},~\left\langle\phi_{B}\phi_{C}\middle|\phi_{A}\phi_{0}\right\rangle$, and $\mathcal{S}$ represent color, flavour, and spin overlap factors, respectively. For the transition of a hybrid into two mesons, $\mathcal{C}=\frac{2}{3}$ is a constant, and $\mathcal{S}$ can be expressed by a $9-j$ symbol as~\cite{Ma:2025cew}
\begin{equation}
\begin{split}
   \mathcal{S} = &\sum_{S} \sqrt{6(2S_B+1)(2S_C+1)(2S_{q\bar{q}}+1)}
   \begin{Bmatrix}
   \frac{1}{2} & \frac{1}{2} & S_B \\
   \frac{1}{2} & \frac{1}{2} & S_C \\
   S_{q\bar{q}} & 1 & S
   \end{Bmatrix}\\
   &\times \left\langle S_{q\bar{q}}, M_{S_{q\bar{q}}}; 1, M_{s_g} \middle| S, M_{S_B} + M_{S_C} \right\rangle\\
   &\times\left\langle S_B, M_{S_B}; S_C, M_{S_C} \middle| S, M_{S_B} + M_{S_C} \right\rangle.
\end{split}
\end{equation}
$\mathcal{I}^{(13)}$ and $\mathcal{I}^{(24)}$ denote to the momentum-space integrals, whose explicit form is~\cite{Ma:2025cew}
\begin{eqnarray}
        \mathcal{I}^{(24)} &=& \int \frac{d^3 \boldsymbol{q} d^3 \boldsymbol{k}}{\sqrt{2 m_g}(2 \pi)^6} \psi_{L_{q\bar{q}} M_{L_{q\bar{q}}}}(\boldsymbol{P}+\boldsymbol{q}+\frac{m_1-m_2}{2m_1+2m_2}\boldsymbol{k})\nonumber\\ &&\times\psi_{L_{q\bar{q},g} M_{L_{q\bar{q},g}}}(\boldsymbol{k}) \psi_{L_B M_{L_B}}^*(\frac{m_4}{m_1+m_4}\boldsymbol{P}+\boldsymbol{q}-\frac{\boldsymbol{k}}{2}) \nonumber\\
        &&\times\psi_{L_C M_{L_C}}^*(\frac{m_4}{m_2+m_4}\boldsymbol{P}+\boldsymbol{q}+\frac{\boldsymbol{k}}{2}),\\
        \mathcal{I}^{(13)} &=& \int \frac{d^3 \boldsymbol{q} d^3 \boldsymbol{k}}{\sqrt{2 m_g}(2 \pi)^6} \psi_{L_{q\bar{q}} M_{L_{q\bar{q}}}}(-\boldsymbol{P}+\boldsymbol{q}+\frac{m_1-m_2}{2m_1+2m_2}\boldsymbol{k})\nonumber\\
        &&\times \psi_{L_{q\bar{q},g} M_{L_{q\bar{q},g}}}(\boldsymbol{k}) \psi_{L_B M_{L_B}}^*(\frac{-m_4}{m_2+m_4}\boldsymbol{P}+\boldsymbol{q}+\frac{\boldsymbol{k}}{2})\nonumber\\
        &&\times\psi_{L_C M_{L_C}}^*(\frac{-m_4}{m_1+m_4}\boldsymbol{P}+\boldsymbol{q}-\frac{\boldsymbol{k}}{2}),
\end{eqnarray}
with $m_i$ being the mass of $i$-th (anti-)quark, $\boldsymbol{P}$ is the momentum of final mesons in the rest frame of hybrid, $\boldsymbol{q}$ is the relative momentum between the created quark and antiquark, and $\boldsymbol{k}$ is the relative momentum between the constituent gluon and the $q\bar{q}$ cluster in the hybrid. Finally, the decay width can be calculated as
\begin{equation}
\begin{split}
   \Gamma_{A\to BC} =& \frac{1}{1+\delta_{BC}} \frac{p_B E_B E_C}{\pi M_A} \\
   &\times\sum_{M_{J_A},M_{J_B},M_{J_C}} \frac{|\mathcal{M}^{M_{J_A},M_{J_B},M_{J_C}}|^2}{2J_A + 1},
\end{split}
\end{equation}
where $E_i=\sqrt{\boldsymbol{P}^2+m_i^2}$ is the energy of the final meson $i$, $M_A$ and $J_A$ are the mass and total angular momentum of the hybrid $A$, and $\delta_{BC}$ is a factor to avoid repeated calculation if $B$ and $C$ are the same.

\section{Numerical results and discussions}
\label{sec3}

\subsection{Mass spectra of $c\bar{c}g$}
As mentioned above, to demonstrate the universality of QCD, the model parameters used in this work, as listed in Table~\ref{tab:placeholder}, are exactly the same as those used in previous meson spectra calculations~\cite{Vijande:2004he,Yang:2011rp,Chen:2024ukv}. These model parameters are obtained from a fit to ground state mesons in the characteristic energy range from the light $q\bar{q}$ to the charmonium $c\bar{c}$ sector, with the included mesons listed in Table~\ref{tab:meson spectrum}.
\renewcommand\tabcolsep{0.6cm}
\textit{}\renewcommand{\arraystretch}{1.7}   
   \begin{table}[!htbp]
       \centering
\caption{Parameters of the constituent quark model used in this work, which is just the same as Ref.~\cite{Vijande:2004he}. }
       \label{tab:placeholder}
       \begin{tabular}{c|cc cc}
 \toprule[1pt]\toprule[1pt]
 Gluon mass      &$m_g$ (MeV)    &~~~~450\\\hline
                   &$m_u$=$m_d$ (MeV)   &~~~~313\\
  Quark masses      &$m_s$ (MeV)  &~~~~555\\
                    &$m_c$ (MeV)  &~~~~1752\\  
                    &$m_b$ (MeV)  &~~~~5100\\\hline

                    &$m_\pi$ (fm$^{-1}$)  &~~~~0.70\\
                    &$m_\sigma$ (fm$^{-1}$)  &~~~~3.42\\
                    &$m_\eta$ (fm$^{-1}$)  &~~~~2.77\\
OBE     &$m_K$ (fm$^{-1}$)  &~~~~2.51\\
  &$\Lambda_\pi=\Lambda_\sigma$ (fm$^{-1}$) &~~~~4.20\\
                     &$\Lambda_\eta=\Lambda_K$ (fm$^{-1}$) &~~~~5.20\\
                     &$g_{ch}^2/(4\pi)$  &~~~~0.54\\
                    &$\theta_p$($\degree$)     &~~~~-15\\ \hline
                    
                    &$a_c$ (MeV)  &~~~~430\\
    Confinement     &$\Delta$ (MeV)  &~~~181.1\\
                    &$\mu_c$ (fm$^{-1}$) &~~~~0.7\\
                    &$a_{s}$     &~~~~0.777\\ \hline
                 &$\alpha_0$  &~~~~2.12\\
                 &$\Lambda_0$ (fm$^{-1}$)  &~~~~0.113\\
        OGE            &$\mu_0$ (MeV)  &~~~~36.976\\
                   &$\hat{r}_0$ (MeV$\cdot$fm)  &~~~~28.17\\
                    &$\hat{r}_g$ (MeV$\cdot$fm)  &~~~~34.5\\
\bottomrule[1pt]\bottomrule[1pt]
\end{tabular}
\end{table}

\renewcommand\tabcolsep{0.1cm}
\renewcommand{\arraystretch}{1.7}
\begin{table}[h]
\centering
\caption{Meson spectrum calculated by using screen parameters and confinement potentials (unit: MeV).}
\label{tab:meson spectrum}
\begin{tabular}{cccccccccc}
\toprule[1pt]\toprule[1pt]
~     &$\pi$       &$\rho$  &$K$     &$K^\star$~          &$\omega$~                                               \\ \hline
CHQM~ &143.50      &775.44    &472.56~          &908.39~             &664.75~     \\
Expt~ &139.57    &775.26      &493.68~          &891.67~             &782.66~        \\ \hline

~     &$\eta$       &$h_1(1170)$  &$a_1(1260)$     &$b_1(1235)$~          &$D_0^{*}(2300)$~                                               \\ \hline
CHQM~ &583.41      &1231.90      &1206.63~          &1236.15~             &2440.64~     \\
Expt~ &547.86      &1166.00      &1230.00~          &1229.50~             &2343.00~        \\ \hline

 ~       &$D$~      &$D^{*}$~  &$D_s$~    &$D_s^{*}$~  &$D_{s0}^{*}(2317)$~ \\ \hline
CHQM~  &1845.22~    &1976.03~              &1922.41~         &2061.76~             &2457.93~        \\
Expt~  &1869.50~    &2006.85~             &1968.35~         &2106.60~             &2317.80~                                                                                             \\  \hline
~     &$\eta_c(1S)$       &$J/\psi(1S)$  &$\chi_{c0}(1P)$     &$\chi_{c1}(1P)$~          &$h_c(1P)$~                                               \\ \hline
CHQM~ &2988.85      &3096.66      &3470.61~          &3516.87~             &3567.70~     \\
Expt~ &2984.10      &3096.90      &3414.71~          &3510.67~             &3525.37~        \\ \bottomrule[1pt]\bottomrule[1pt]
\end{tabular}
\end{table}

In the charm sector, there will be no Goldstone boson exchanges any more, so that there will only have color confinement and gluon exchanges, which is just the same as the ordinary quark model. Since one gluon exchange potential can be directly derived from the $t$-channel quark-(anti)quark scattering amplitude, the main difference just comes from the choice of the phenomenological color confinement. In this work, to use the same model parameter of Ref.~\cite{Vijande:2004he} to exhibit the unity of QCD, we also use the screened form of color confinement. If comparing with the classical linear confinement within the meson system, our string tension will be $\left(\boldsymbol{\lambda}_i^c \cdot \boldsymbol{\lambda}_j^{c*} \right)a_c\mu_c=0.315~\mathrm{GeV^2}$, which is obtained from the screened confinement potential,
\begin{eqnarray}
   V_{q\bar{q}}^{\mathrm{CON},C}(\boldsymbol{r}_{ij}) &=&
   \left( -\boldsymbol{\lambda}_i^c \cdot \boldsymbol{\lambda}_j^{c*} \right)
   \left[ -a_c \left( 1 - e^{-\mu_c r_{ij}} \right) + \Delta \right].
\end{eqnarray}
Here, $\left( \boldsymbol{\lambda}_i^c \cdot \boldsymbol{\lambda}_j^{c*} \right)=16/3$, $a_c=0.43$~GeV and $\mu_c=0.7~\mathrm{fm^{-1}}$. Our string tension is larger than the most linear string tensions adopted in other approaches~\cite{Qian:2026inl,Berwein:2015vca,Farina:2020slb,Akbar:2024jda,Lebed:2017xih,Kleiv:2014kua,Chen:2013zia,Sultan:2014oua,Chanowitz:1982qj,Barnes:1995hc,Kalashnikova:2016bta,Oncala:2017hop,Buisseret:2006wc}, which typically lie in the range $0.11-0.21$ GeV$^2$~\cite{Barnes:1981ac}. Such difference arises because our confinement potential is of the screened form rather than a pure linear potential, which is linear only for \(r\ll 1/\mu_c\simeq 1.4\) fm and saturates at larger \(r\). Also, our confinement contains an additional mass shift parameter $\Delta$. Thus, our screened confinement may not be directly compared with the string tension of the pure linear potential. As for the strong coupling constant, our $\alpha_s(c\bar{c})=0.288$ just fall within the common range $0.25-0.41$ supported by lattice QCD, pNRQCD, and various phenomenological models ~\cite{Qian:2026inl,Berwein:2015vca,Lebed:2017xih,Kleiv:2014kua,Chen:2013zia,Barnes:1995hc,Kalashnikova:2016bta,Oncala:2017hop,Buisseret:2006wc}, although a subset of constituent gluon and potential model calculations employ larger fixed $\alpha_S$ values around $0.5-0.6$~\cite{Farina:2020slb,Sultan:2014oua,Barnes:1981ac}. Nevertheless, although there exists some differences between our model with previous works, since both our spectra on the low-lying traditional mesons are not far away from the previous models and experiments, we think that our screened confinement is reasonable.

From the comparison in Table~\ref{tab:meson spectrum}, it is evident that the fit results are in good agreement with the experimental data for most mesons. Since our model introduces only one new parameter, namely the constituent-gluon mass, and given the universal running strong coupling constant in Eq.~(\ref{eq:runalphas}), we directly extend these model parameters from the meson sector to the hybrid sector, obtaining the mass spectra of ground hybrids with various quantum numbers given in Table~\ref{tab:ccbarg mass spectrum}. 

Throughout the calculation, we fix $J_g=1$ and categorize the spectra into two main cases: one with $L_{q\bar{q},g}=0$, and the other with $L_{q\bar{q},g}=1$. From the perspective of the three-body problem, this distinction merely reflects whether there is an orbital excitation between the $q\bar{q}$ cluster and the constituent gluon. However, considering the constraint $J_g=1$ together with Ref.~\cite{Qian:2026inl}, we interpret this difference in $L_{q\bar{q},g}$ as describing how the gluon manifests as a constituent. For $L_{q\bar{q},g}=0$, the constituent gluon couples directly to the $q\bar{q}$ cluster, meaning that the gauge field with quantum number $1^{--}$ directly serves as the constituent. For $L_{q\bar{q},g}=1$, the gluon can serve as a constituent only after being excited with an orbital angular momentum $L_{q\bar{q},g}=1$. In this case, the quantum number of the constituent gluon is $1^{+-}$, and such an excitation of the gauge field, as shown in Ref.~\cite{Qian:2026inl}, can be directly related to the commonly used concept of a ``transverse electric'' (TE) mode constituent gluon. In fact, compared to the direct gauge-field description, the excited $1^{+-}$ constituent gluon is more commonly adopted in various works~\cite{Qian:2026inl,Cheung:2016bym,Farina:2020slb,HadronSpectrum:2012gic,Wang:2025clb,Akbar:2024jda,Vairo:2017dmh,Lebed:2017xih,TarrusCastella:2017lex,TarrusCastella:2015mbn,Berwein:2015vca,Braaten:2014qka,Kleiv:2014kua,Sultan:2014oua,Barnes:1982tx,Chanowitz:1982qj,Barnes:1995hc,Kalashnikova:2016bta,Agaev:2025llz,Chen:2013zia,Oncala:2017hop}, especially lattice QCD~\cite{Cheung:2016bym,HadronSpectrum:2012gic,TarrusCastella:2017lex,TarrusCastella:2015mbn,Berwein:2015vca,Braaten:2014qka}, as the explicit constituent in hybrid studies. Nevertheless, a few works are based on the $1^{--}$ constituent gluon ansatz~\cite{Iddir:2007dq,Iddir:2006sh}. Thus, for completeness and to adhere to the spirit of the three-body treatment, we perform a systematic study of both cases in this work.

\renewcommand\tabcolsep{0.08cm}
\renewcommand{\arraystretch}{1.7}
\begin{table}[!htbp]
		\centering
\caption{Mass spectra $(M_{\text{min}}, M_{\text{max}})$ of ground charmonium hybrid states obtained by varying the mass of the $1^{--}$ constituent gluon from 450 MeV to 950 MeV. All masses are in units of MeV. The previous theoretical results from other models are listed in the last column for comparison~\cite{Qian:2026inl,Cheung:2016bym,Farina:2020slb,HadronSpectrum:2012gic,Wang:2025clb,Akbar:2024jda,Vairo:2017dmh,Lebed:2017xih,TarrusCastella:2017lex,TarrusCastella:2015mbn,Berwein:2015vca,Braaten:2014qka,Kleiv:2014kua,Sultan:2014oua,Barnes:1982tx,Chanowitz:1982qj,Barnes:1995hc,Kalashnikova:2016bta,Agaev:2025llz,Chen:2013zia,Oncala:2017hop,Iddir:2007dq,Iddir:2006sh} and include potential models \cite{Qian:2026inl,Farina:2020slb,Akbar:2024jda,Sultan:2014oua,Barnes:1982tx,Chanowitz:1982qj,Iddir:2007dq}, lattice QCD \cite{Cheung:2016bym,HadronSpectrum:2012gic,TarrusCastella:2017lex,TarrusCastella:2015mbn,Berwein:2015vca,Braaten:2014qka}, QCD sum rules \cite{Kleiv:2014kua,Barnes:1982tx,Chen:2013zia}, and the Born-Oppenheimer approximation, often implemented within effective field theories~\cite{Vairo:2017dmh,Lebed:2017xih,TarrusCastella:2017lex,TarrusCastella:2015mbn,Berwein:2015vca,Braaten:2014qka,Oncala:2017hop}.} \label{tab:ccbarg mass spectrum}
\begin{tabular}{cccl}
\toprule[1pt]\toprule[1pt]
$J^{\mathrm{PC}}$ &$(S_{q\bar{q}}, L_{q\bar{q}}, L_{q\bar{q},g}, L_{g}, L, J)$ & $(M_{\text{min}}, M_{\text{max}})$  &Others\\
			\midrule[1pt]
$1^{+-}$  &(0, 0, 0, 1, 0, 1) & (3843,4135)  & -\\
$0^{++}$     &(1, 0, 0, 0, 0, 0) & (3732,4015)  & -\\
$1^{++}$	 &(1, 0, 0, 1, 0, 1) &(3795,4086)  & -\\
$2^{++}$	 &(1, 0, 0, 1, 0, 2) &(3881,4172)  & -\\
            \midrule[1pt]
$0^{-+}$ &(0, 1, 0, 0, 1, 0) & (3944,4277)  &4090~\cite{Iddir:2007dq,Iddir:2006sh}\\
$1^{-+}$	  &(0, 1, 0, 1, 1, 1) & (3950,4285)  &4090~\cite{Iddir:2007dq,Iddir:2006sh}\\
$2^{-+}$	  &(0, 1, 0, 2, 1, 2) & (3961,4301)  &-\\
$1^{--}$      &(1, 1, 0, 0, 1, 1) & (3950,4281)  &4090~\cite{Iddir:2007dq,Iddir:2006sh}\\
$0^{--}$	&(1, 1, 0, 1, 1, 0) & (3925,4244)  & 4090~\cite{Iddir:2007dq,Iddir:2006sh} \\
$1^{--}$	&(1, 1, 0, 1, 1, 1) & (3921,4261)  &-\\
$2^{--}$	&(1, 1, 0, 1, 1, 2) & (3985,4316)  &-\\
$1^{--}$	&(1, 1, 0, 2, 1, 1) & (3902,4254)  &-\\
$2^{--}$	&(1, 1, 0, 2, 1, 2) & (3927,4274)  &-\\
$3^{--}$	&(1, 1, 0, 2, 1, 3) & (3987,4323)  &-\\
            \midrule[1.pt]
$1^{--}$  &(0, 0, 1, 1, 1, 1) & (4225,4491)  &4150$\sim$4467~\cite{Qian:2026inl,HadronSpectrum:2012gic,Wang:2025clb,Akbar:2024jda,Sultan:2014oua,Vairo:2017dmh,Lebed:2017xih,TarrusCastella:2017lex,TarrusCastella:2015mbn,Kleiv:2014kua,Berwein:2015vca,Braaten:2014qka,Kalashnikova:2016bta,Farina:2020slb,Cheung:2016bym,Barnes:1982tx,Chanowitz:1982qj,Barnes:1995hc,Oncala:2017hop}\\
$0^{-+}$	&(1, 0, 1, 1, 1, 0) & (4139,4437)  &4126$\sim$4510~\cite{Qian:2026inl,HadronSpectrum:2012gic,Wang:2025clb,Akbar:2024jda,Sultan:2014oua,Vairo:2017dmh,Lebed:2017xih,TarrusCastella:2017lex,TarrusCastella:2015mbn,Kleiv:2014kua,Berwein:2015vca,Braaten:2014qka,Kalashnikova:2016bta,Farina:2020slb,Cheung:2016bym,Barnes:1982tx,Chanowitz:1982qj,Barnes:1995hc,Oncala:2017hop}\\
$1^{-+}$	&(1, 0, 1, 1, 1, 1) & (4219,4470) &4148$\sim$4510~\cite{Qian:2026inl,HadronSpectrum:2012gic,Wang:2025clb,Akbar:2024jda,Sultan:2014oua,Vairo:2017dmh,Lebed:2017xih,TarrusCastella:2017lex,TarrusCastella:2015mbn,Kleiv:2014kua,Berwein:2015vca,Braaten:2014qka,Kalashnikova:2016bta,Farina:2020slb,Cheung:2016bym,Barnes:1982tx,Chanowitz:1982qj,Barnes:1995hc}\\
$2^{-+}$	&(1, 0, 1, 1, 1, 2) & (4241,4519)  &4040$\sim$4510~\cite{Qian:2026inl,HadronSpectrum:2012gic,Wang:2025clb,Akbar:2024jda,Sultan:2014oua,Vairo:2017dmh,Lebed:2017xih,TarrusCastella:2017lex,TarrusCastella:2015mbn,Agaev:2025llz,Kleiv:2014kua,Berwein:2015vca,Braaten:2014qka,Kalashnikova:2016bta,Farina:2020slb,Cheung:2016bym,Barnes:1982tx,Chanowitz:1982qj,Barnes:1995hc,Chen:2013zia}\\
        \midrule[1pt]
$0^{++}$  &(0, 1, 1, 0, 2, 0) & (4310,4634)  &4299$\sim$4690~\cite{HadronSpectrum:2012gic,Wang:2025clb,Akbar:2024jda,Vairo:2017dmh,Lebed:2017xih,TarrusCastella:2017lex,TarrusCastella:2015mbn,Kleiv:2014kua,Berwein:2015vca,Braaten:2014qka,Farina:2020slb,Cheung:2016bym}\\
$1^{++}$	&(0, 1, 1, 1, 2, 1) & (4253,4548)  &4090$\sim$4490~\cite{HadronSpectrum:2012gic,Wang:2025clb,Akbar:2024jda,Sultan:2014oua,Vairo:2017dmh,Lebed:2017xih,TarrusCastella:2017lex,TarrusCastella:2015mbn,Kleiv:2014kua,Berwein:2015vca,Braaten:2014qka,Farina:2020slb}\\
$2^{++}$	&(0, 1, 1, 2, 2, 2) & (4297,4601)  &4170$\sim$4500~\cite{HadronSpectrum:2012gic,Wang:2025clb,Akbar:2024jda,Vairo:2017dmh,Lebed:2017xih,TarrusCastella:2017lex,TarrusCastella:2015mbn,Agaev:2025llz,Kleiv:2014kua,Berwein:2015vca,Braaten:2014qka,Farina:2020slb,Cheung:2016bym}\\
$1^{+-}$	&(1, 1, 1, 0, 2, 1) & (4313,4636)  &4090$\sim$4690~\cite{HadronSpectrum:2012gic,Wang:2025clb,Akbar:2024jda,Sultan:2014oua,Vairo:2017dmh,Lebed:2017xih,TarrusCastella:2017lex,TarrusCastella:2015mbn,Berwein:2015vca,Braaten:2014qka,Farina:2020slb}\\
$0^{+-}$	&(1, 1, 1, 1, 2, 0) & (4219,4516)  &4229$\sim$4670~\cite{Sultan:2014oua,HadronSpectrum:2012gic,Wang:2025clb,Akbar:2024jda,Vairo:2017dmh,Lebed:2017xih,TarrusCastella:2017lex,TarrusCastella:2015mbn,Kleiv:2014kua,Berwein:2015vca,Braaten:2014qka,Farina:2020slb,Cheung:2016bym}\\
$1^{+-}$	&(1, 1, 1, 1, 2, 1) & (4261,4564)  &4270$\sim$4670~\cite{HadronSpectrum:2012gic,Wang:2025clb,Akbar:2024jda,Vairo:2017dmh,Lebed:2017xih,TarrusCastella:2017lex,TarrusCastella:2015mbn,Berwein:2015vca,Braaten:2014qka,Farina:2020slb,Cheung:2016bym}\\
$2^{+-}$	&(1, 1, 1, 1, 2, 2) & (4280,4582)  &4273$\sim$4670~\cite{HadronSpectrum:2012gic,Wang:2025clb,Akbar:2024jda,Sultan:2014oua,Vairo:2017dmh,Lebed:2017xih,TarrusCastella:2017lex,TarrusCastella:2015mbn,Berwein:2015vca,Braaten:2014qka,Farina:2020slb,Cheung:2016bym}\\
$1^{+-}$	&(1, 1, 1, 2, 2, 1) & (4310,4626)  &4170$\sim$4530~\cite{HadronSpectrum:2012gic,Wang:2025clb,Akbar:2024jda,Vairo:2017dmh,Lebed:2017xih,TarrusCastella:2017lex,TarrusCastella:2015mbn,Berwein:2015vca,Braaten:2014qka,Farina:2020slb,Cheung:2016bym}\\
$2^{+-}$	&(1, 1, 1, 2, 2, 2) & (4315,4623)  &4273$\sim$4550~\cite{HadronSpectrum:2012gic,Wang:2025clb,Akbar:2024jda,Vairo:2017dmh,Lebed:2017xih,TarrusCastella:2017lex,TarrusCastella:2015mbn,Berwein:2015vca,Braaten:2014qka,Farina:2020slb,Cheung:2016bym}\\
$3^{+-}$	&(1, 1, 1, 2, 2, 3) & (4324,4622)  &4170$\sim$4636~\cite{HadronSpectrum:2012gic,Wang:2025clb,Akbar:2024jda,Vairo:2017dmh,Lebed:2017xih,TarrusCastella:2017lex,TarrusCastella:2015mbn,Berwein:2015vca,Braaten:2014qka,Farina:2020slb,Cheung:2016bym}\\
\bottomrule[1pt]\bottomrule[1pt]
		\end{tabular}
\end{table}

As shown in Table~\ref{tab:ccbarg mass spectrum}, using the same model parameters from the meson fit, together with the single new parameter $m_g=450$--$950$ MeV, our ground hybrid spectra with various quantum numbers are in good agreement with the corresponding results from other works~\cite{Qian:2026inl,Cheung:2016bym,Farina:2020slb,HadronSpectrum:2012gic,Wang:2025clb,Akbar:2024jda,Vairo:2017dmh,Lebed:2017xih,TarrusCastella:2017lex,TarrusCastella:2015mbn,Berwein:2015vca,Braaten:2014qka,Kleiv:2014kua,Sultan:2014oua,Barnes:1982tx,Chanowitz:1982qj,Barnes:1995hc,Kalashnikova:2016bta,Agaev:2025llz,Chen:2013zia,Oncala:2017hop,Iddir:2007dq,Iddir:2006sh}. In particular, taking $m_g=450$~MeV, we find that most of our mass spectra for these $c\bar{c}g$ hybrids already match the results of existing works. Furthermore, for the $L_{q\bar{q},g}=1$ excited-gluon case, we reproduce the mass ordering among the $0^{-+}$, $1^{\pm-}$, and $2^{-+}$ $c\bar{c}g$ hybrids, as indicated by Refs.~\cite{Qian:2026inl,Barnes:1982tx,Chanowitz:1982qj}:
\begin{eqnarray}
    m_{0^{-+}}<m_{1^{-+}}<m_{1^{--}}<m_{2^{-+}} .
\end{eqnarray}
These consistencies therefore demonstrate that our constituent gluon model with $m_g=450$~MeV can be applied not only to $1^{-+}$ light hybrid mesons but also to charmonium hybrids with various quantum numbers, which, in our view, deeply reflects the underlying unification of QCD.

Given the success of our systematic reproduction of the ground $c\bar{c}g$ masses, we now examine the spectra from a global perspective. A notable feature emerges: our spectra are dense within two narrow energy ranges, one around 3.9~GeV and the other around 4.2~GeV, depending on whether the gluon is excited. Notably, this behavior is similar to that observed in baryon spectra calculations, where the coupling between spin and orbital angular momenta of the three constituents leads to a series of multiplets with similar masses.

In summary, within the framework of the constituent gluon model, we present comprehensive results for the mass spectra of charmonium hybrid states with various quantum numbers. In particular, by introducing only one new parameter, $m_g=450$~MeV, as the constituent gluon mass, we reproduce most mass positions obtained by other works, demonstrating the effectiveness of our model and the unity of QCD. Moreover, when the constituent gluon mass is fixed, the masses of the charmonium hybrids are strongly dependent on the quantum number assignments among the quark, antiquark, and gluon constituents. In particular, although the predicted hybrid masses vary with the constituent gluon mass, the mass splittings between different quantum-number configurations remain relatively stable. This is a crucial feature for future experimental verification: if our model is correct, the identification of one charmonium hybrid would imply the existence of others at predictable mass splittings.

\subsection{Decay properties}

To investigate the potential for these $c\bar{c}g$ hybrids to be experimentally verified, we perform systematic calculations of their two-body decay widths. The decay mechanism in Fig.~\ref{fig:decay_diagram} indicates that all these $c\bar{c}g$ hybrids should mainly decay into pairs of open-charm or charm-strange mesons. Using the wave functions obtained from the mass spectra calculations, we calculate all allowed partial decay widths of the corresponding charmonium hybrid mesons, with the results collected in Tables~\ref{tab:ccgTES}--\ref{tab:ccgP}.

\renewcommand\tabcolsep{0.25cm}
\renewcommand{\arraystretch}{1.7}
\begin{table}[!htbp]
\centering
\caption{Decay widths $(\Gamma_{\text{min}},\Gamma_{\text{max}})$ of the $c\bar{c}g$ hybrids obtained by varying the mass of the $1^{--}$ constituent gluon from 450 to 950 MeV. All widths are in MeV. The constituent gluon is excited with $L_{q\bar{q},g}=1$, and there is no excitation between $c\bar{c}$. Here, ``$\times$'' indicates that the decay channel is kinematically forbidden, ``0'' indicates that the mode cannot occur due to selection rules, while ``$\approx$0'' means the decay width is smaller than 0.01~MeV.}
\label{tab:ccgTES}
\footnotesize
\begin{tabular}{l|ccccc}
\toprule[1pt]\toprule[1pt]
\diagbox{Modes}{$J^{PC}$}
 & $1^{--}$ & $0^{-+}$ & $1^{-+}$ & $2^{-+}$   \\
\midrule[1pt]
$D\bar{D}$  & 0 & 0 & 0 & 0 \\
$D\bar{D}^*$ & $\approx0$ & $\approx0$ & $\approx0$ & $\approx0$   \\
$D\bar{D}^*_0(2300)$ & $(\times, \approx0)$ & $(\times, 25.1)$ & $(\times, \approx0)$ & $(\times, \approx0)$   \\
$D^*\bar{D}^*$ & 0 & 0 & 0 & 0   \\
$D\bar{D}^*_2(2460)$  & $(\times, \approx0)$ & $(\times, \approx0)$ & $(\times, \approx0)$ & $(\times, 9.4)$   \\
$D\bar{D}_1(2430)$  & $(\times, 9.7)$ & $(\times, 0)$ & $(\times, 9.9)$ & $(\times, \approx0)$  \\
$D\bar{D}_1(2420)$ & $(\times, 9.1)$ & $(\times, 0)$ & $(\times, 9.3)$ & $(\times, \approx0)$    \\
\hline
$D_s\bar{D}_s$ & 0 & 0 & 0 & 0   \\
$D_s\bar{D}^*_s$  & $\approx0$ & $(\approx0, 0.1)$ & $\approx0$ & $\approx0$  \\
$D_s\bar{D}^*_{s0}(2317)$ &$(\times, \approx0)$ & $(\times, 35.6)$ & $(\times, 0)$ & $(\times, \approx0)$  \\
$D_s^*\bar{D}^*_s$  & 0 & 0 & 0 & 0   \\
\midrule[1pt]
Total width & $(0, 18.8)$ & $(0, 60.7)$ & $(0, 19.2)$ & $(0, 9.4)$   \\
\midrule[1pt]
Ref.~\cite{Farina:2020slb} & $\times\sim41$ & $\times\sim110$ & $\times\sim28$ & $\times\sim20$\\
\midrule[1pt]
Ref.~\cite{Qian:2026inl} & $0\sim60$ &  &  & $0\sim30$\\
\bottomrule[1pt]\bottomrule[1pt]
\end{tabular}
\end{table}

\renewcommand\tabcolsep{0.24cm}
\renewcommand{\arraystretch}{1.7}
\begin{table*}[!htbp]
\centering
\caption{Decay widths $(\Gamma_{\text{min}},\Gamma_{\text{max}})$ of the $c\bar{c}g$ hybrids obtained by varying the mass of the $1^{--}$ constituent gluon from 450 to 950 MeV. The constituent gluon is excited with $L_{q\bar{q},g}=1$, and $c\bar{c}$ is excited with $L_{q\bar{q}}=1$. Here, ``$\times$'' indicates that the decay channel is kinematically forbidden, ``0'' indicates that the mode cannot occur due to selection rules, while ``$\approx$0'' means the decay width is smaller than 0.01~MeV.} \label{tab:ccgTEP}
\begin{tabular}{l|cccccccccc}
\toprule[1pt]\toprule[1pt]
\diagbox[width=2.2cm,height=1.0cm,trim=l]{Modes}{$J^{PC}$}
 & $0^{++}$ & $1^{++}$ & $2^{++}$ & $1^{+-}$ & $0^{+-}$ & $1^{+-}$ & $2^{+-}$ & $1^{+-}$ & $2^{+-}$ & $3^{+-}$  \\
\midrule[1pt]
$D\bar{D}$  & 0 & 0 & 0 & 0 &0 & 0 & 0 & 0 & 0 & 0  \\
$D\bar{D}^*$  & 0.0 & $\approx$0 & 0 & 0 & 0 &$\approx$0 & $\approx$0 & $\approx$0 & $\approx$0 & $\approx$0  \\
$D\bar{D}^*_0(2300)$ & 0 & $(\times, 0)$ & 0 & $(\approx0, 1.5)$ & ($\times$, 1.8) & $(\times$, 5.5) & ($\times$,0) & (5.9, 19.6) & 0 & $\approx$0  \\
$D^*\bar{D}^*$ & 0 & 0 & 0 & 0 & 0 & 0 & 0 & 0 & 0 & 0  \\
$D^*\bar{D}^*_0(2300)$  & ($\times$, 3.8) & ($\times$, 7.1) & ($\times$, 5.1) & ($\times$, $\approx$0) & ($\times$, 0.4) & ($\times$, $\approx$0) & ($\times$, $\approx$0) & ($\times$, $\approx$0) & ($\times$, 0.2) & ($\times$, 0)  \\
$D\bar{D}^*_2(2460)$  & ($\times$, 0) & ($\times$, 0.3) & ($\times$, 0.1) & ($\times$, 1.5) & ($\times$, 0) & ($\times$, 2.7) & ($\times$, 3.3) & ($\times$, 0.4) & ($\times$, 0.7) & ($\times$, 6.8)   \\
$D\bar{D}_1(2430)$  & ($\times$, 5.1) & ($\times$, 9.4) & ($\times$, 7.8) & ($\times$, 1.0) & ($\times$, 1.5) & ($\times$, 1.1) & ($\times$, 1.2) & ($\times$, 4.5) & ($\times$, 3.8) & ($\times$, $\approx$0)  \\
$D\bar{D}_1(2420)$ & ($\times$, 5.0) & ($\times$, 9.2) & ($\times$, 8.0) & ($\times$, 1.0) & ($\times$, 3.1) & ($\times$, 1.3) & ($\times$, 1.3) & ($\times$, 4.7) & ($\times$, 3.8) & ($\times$, $\approx$0)  \\
\hline
$D_s\bar{D}_s$  & 0 & 0 & 0 & 0 & 0 & 0 & 0 & 0 & 0 & 0  \\
$D_s\bar{D}^*_s$  & 0 & $\approx$0 & 0 & 0 & 0 &$\approx$0 & $\approx$0 & $\approx$0 & $\approx$0 & $\approx$0  \\
$D_s\bar{D}^*_{s0}$ & ($\times$, 0) & ($\times$, 0) & ($\times$, 0) & ($\times$, 1.0) & ($\times$, 0) & ($\times$, 4.0) & ($\times$, 0) & ($\times$, 6.1) & ($\times$, 0) & ($\times$, $\approx$0)  \\
$D_s^*\bar{D}^*_s$  & 0 & 0 & 0 & 0 & 0 & 0& 0 & 0 & 0 & 0  \\
$D_s^*\bar{D}^*_{s0}$ & ($\times$, 2.4) & ($\times$, 1.0) & ($\times$, 2.4) & ($\times$, $\approx$0) & $\times$ & ($\times$, $\approx$0) & ($\times$, $\approx$0) & $\times$ & $\times$ & ($\times$, 0)   \\
\midrule[1pt]
Total width & (0, 16.3) & (0, 27) & (0, 23.4) & (0, 6.0) & (0, 6.8) & (0, 14.6) & (0, 5.8) & (5.9, 35.3) & (0, 8.5) & (0, 6.8)  \\
\midrule[1pt]
Ref.~\cite{Farina:2020slb} & 29.48 & 18.34 & 35.50 & 28.39 & 28.52 & 21.39 & 34.54 & 80.88 & 65.81 & 9.10  \\
\bottomrule[1pt]\bottomrule[1pt]
\end{tabular}
\end{table*}

\renewcommand\tabcolsep{0.2cm}
\renewcommand{\arraystretch}{1.7}
\begin{table}[!htbp]
\centering
\caption{Decay widths $(\Gamma_{\text{min}},\Gamma_{\text{max}})$ of the $c\bar{c}g$ hybrids obtained by varying the mass of the $1^{--}$ constituent gluon from 450 to 950 MeV. The constituent gluon is {\it not} excited, i.e., $L_{q\bar{q},g}=0$, and there is no excitation between $c\bar{c}$. Here, ``$\times$'' indicates that the decay channel is kinematically forbidden, ``0'' indicates that the mode cannot occur due to selection rules, while ``$\approx$0'' means the decay width is smaller than 0.01~MeV.}
\label{tab:ccgS}
\footnotesize
\begin{tabular}{l|ccccc}
\toprule[1pt]\toprule[1pt]
\diagbox{Modes}{$J^{PC}$}
 & $1^{+-}$ & $0^{++}$ & $1^{++}$ & $2^{++}$   \\
\midrule[1pt]
$D\bar{D}$  & 0 & (75.6, 178.7) & 0 & 0 \\
$D\bar{D}^*$ & (16.9, 47.8) & ($\times$, 0) & ($\times$, 91.7) & 0   \\
$D\bar{D}^*_0(2300)$ &0 & 0 &  0 &  0   \\
$D^*\bar{D}^*$ & ($\times$, 52.1) & ($\times$, 29.7) & ($\times$, 0) & ($\times$, 0)   \\
\hline
$D_s\bar{D}_s$  & ($\times$, 0) & ($\times$, 118.4) & ($\times$, 0) & ($\times$, 0)   \\
$D_s\bar{D}^*_s$  & ($\times$, 33.9) & ($\times$, 0) & ($\times$, 71.7) & ($\times$, 0)  \\
$D_s^*\bar{D}^*_s$  & ($\times$, 36.9) & $\times$ & $\times$ & ($\times$, 0)   \\
\midrule[1pt]
Total width & (16.9, 170.7) & (75.6, 326.8) & (0, 163.4) &  0   \\
\bottomrule[1pt]\bottomrule[1pt]
\end{tabular}
\end{table}

\renewcommand\tabcolsep{0.1cm}
\renewcommand{\arraystretch}{1.7}
\begin{table*}[!htbp]
\centering
\caption{Decay widths $(\Gamma_{\text{min}},\Gamma_{\text{max}})$ of the $c\bar{c}g$ hybrids obtained by varying the mass of the $1^{--}$ constituent gluon from 450 to 950 MeV. The constituent gluon is {\it not} excited, i.e., $L_{q\bar{q},g}=0$, and $c\bar{c}$ is excited with $L_{q\bar{q}}=1$. Here, ``$\times$'' indicates that the decay channel is kinematically forbidden, ``0'' indicates that the mode cannot occur due to selection rules, while ``$\approx$0'' means the decay width is smaller than 0.01~MeV.} \label{tab:ccgP}
\begin{tabular}{l|cccccccccc}
\toprule[1pt]\toprule[1pt]
\diagbox[width=2.2cm,height=1.0cm,trim=l]{Modes}{$J^{PC}$}
 & $0^{-+}$ & $1^{-+}$ & $2^{-+}$ & $1^{--}$ & $0^{--}$ & $1^{--}$ & $2^{--}$ & $1^{--}$ & $2^{--}$ & $3^{--}$  \\
\midrule[1pt]
$D\bar{D}$  & 0 & 0 & 0 & (2.4, 6.9) & 0 & (13.2, 35.3) & 0 & (17.5, 45.0) & 0 & 0 \\
$D\bar{D}^*$ & (9.1, 18.2) & (8.6, 18.5) & (7.6, 18.9) & (6.6, 13.9) & (22.0, 35.5) & (8.6, 14.6) & (4.5, 12.6) & (11.1, 18.4) & (18.8, 33.9) & 0  \\
$D\bar{D}^*_0(2300)$ & ($\times$, 0) & ($\times$, 0) & ($\times$, $\approx$0) & ($\times$, 0) & ($\times$, 0) & ($\times$, 0) & ($\times$, $\approx$0) & ($\times$, 0) & ($\times$, 0) & ($\times$, 0)  \\
$D^*\bar{D}^*$ & ($\times$, 31.5) & ($\times$, 32.3) & ($\times$, 33.2) & ($\times$, 2.0) & ($\times$, 0) & ($\times$, 8.5) & 0 & ($\times$, 9.2) & ($\times$, 0) & 0  \\
\hline
$D_s\bar{D}_s$    & 0 & 0 & 0 & (2.3, 3.7) & 0 & (8.2, 19.5) & 0 & (7.3, 24.9) & 0 & 0  \\
$D_s\bar{D}^*_s$  & ($\times$, 11.2) & ($\times$, 11.0) & ($\times$, 10.6) & ($\times$, 9.4) & ($\times$, 27.8) & ($\times$, 11.5) & ($\approx$0, 7.2) & ($\times$, 14.3) & ($\times$, 26.1) & ($\times$, 0)  \\
$D_s^*\bar{D}^*_s$  & ($\times$, 18.7) & ($\times$, 19.3) & ($\times$, 20.2) & ($\times$, 1.6) & ($\times$, 0) & ($\times$, 6.7) & ($\times$, 0) & ($\times$, 7.8) & ($\times$, 0) & 0  \\
\midrule[1pt]
Total width & (9.1, 79.6) & (8.6, 81.1) & (7.6, 82.9) & (11.3, 37.5) & (22.0, 63.6) & (30.0, 96.1) & (4.5, 19.8) & (35.9, 119.6) & (18.8, 60.0) &  0  \\
\bottomrule[1pt]\bottomrule[1pt]
\end{tabular}
\end{table*}

\subsubsection{$c\bar{c}g$ states with $L_{q\bar{q}}=0$ and $L_{q\bar{q},g}=1$}

We begin with the results in Table~\ref{tab:ccgTES}, where the constituent gluon is excited with $L_{q\bar{q},g}=1$ and there is no $c\bar{c}$ excitation. We observe that although our total decay widths differ slightly from those in Refs.~\cite{Qian:2026inl,Farina:2020slb}, the ratios among these widths are very similar. In fact, the ratios reported by Refs.~\cite{Qian:2026inl,Farina:2020slb} are consistent with each other, since both their physical descriptions of the hybrid and their expressions for the decay amplitude are the same; the differences in absolute widths arise from different model parameters. For instance, $m_g$ is taken as 0.6~GeV in one of the analyses but 1.05~GeV in the other. Regarding the comparison between our model and Refs.~\cite{Qian:2026inl,Farina:2020slb}, in addition to the different physical descriptions of the hybrid and different expressions for the decay amplitude, our model parameters are taken directly from meson spectra calculations owing to the assumed unity of QCD. Consequently, the $\alpha_s$ entering the decay amplitude is determined self-consistently and differs between open-charm and open-charm-strange channels, whereas Refs.~\cite{Qian:2026inl,Farina:2020slb} adopt a typical value $\alpha_s \approx 0.5$--$0.6$. Thus, despite the aforementioned differences, the agreement between our results and those of Refs.~\cite{Qian:2026inl,Farina:2020slb} for both the mass spectra and decay widths of these four charmonium hybrid mesons reinforces our confidence in the validity of our model.

Combining these decay results with the spectra in Table~\ref{tab:ccbarg mass spectrum}, we find that with $m_g=450$~MeV, although these four charmonium hybrid mesons have similar masses around 4.2~GeV, all but the $0^{-+}$ state are very narrow. Furthermore, due to the selection rule discussed in Ref.~\cite{Qian:2026inl}, none of these states can decay into pairs of $S$-wave charm(-strange) mesons. Therefore, for experimental searches, the $1^{-+}$ state with exotic quantum numbers is the most promising candidate, and experiments should focus on the $D\bar{D}_1$ channels. For the three states with ordinary quantum numbers, the $0^{-+}$ and $2^{-+}$ states are particularly recommended for experimental verification, as their dominant decay channels are exclusively $D\bar{D}^\ast_0$ and $D_s\bar{D}^\ast_{s0}$, making them easily distinguishable. As for the $1^{--}$ state, since its mass lies around 4.2--4.5~GeV, it is natural to imagine that it may mix with nearby vector charmonia, though only through open-charm(-strange) channels involving at least one orbitally excited charm(-strange) meson. For example, with $m_g=450$~MeV, the mass of this $1^{--}$ hybrid is located around that of the $\psi(4230)$~\cite{ParticleDataGroup:2024cfk}. Thus, within the unquenched quark model, the $\psi(4230)$ may contain a substantial $c\bar{c}g$ component, as suggested by Ref.~\cite{Wang:2025clb}. As for verifying the effects of such a $c\bar{c}g$ component on the $c\bar{c}$ core, this will require further studies of $c\bar{c}$--$c\bar{c}g$ mixing.

\subsubsection{$c\bar{c}g$ states with $L_{q\bar{q}}=1$ and $L_{q\bar{q},g}=1$}

For hybrids with both orbital $c\bar{c}$ excitation and gluon excitation, i.e., $(L_{q\bar{q}},L_{q\bar{q},g})=(1,1)$, these form a group of ten states whose masses are close to one another, as shown in Table~\ref{tab:ccbarg mass spectrum}. Their decay properties, listed in Table~\ref{tab:ccgTEP}, show that they still dominantly decay into channels containing at least one orbitally excited charm(-strange) meson. However, in this case, our total decay widths do not agree with Ref.~\cite{Farina:2020slb} in both absolute values and the relative ratios among different states, which in our view arises mainly from differences in the mass spectra. Comparing our mass spectra with Ref.~\cite{Farina:2020slb}, we observe that although their results are similar to ours for the $(0,1)$ configuration, for the $(1,1)$ configuration the lower mass bounds in their work are much higher than ours. This opens many two-body thresholds and consequently the lower bounds of their total widths do not vanish. Furthermore, their mass splittings among different states differ from ours and are less stable; mass inversions even occur for some states as model parameters vary.

For experimental searches, we again prioritize states with exotic quantum numbers. However, apart from the $0^{+-}$ state, there are two $c\bar{c}g$ states with the same quantum number $2^{+-}$, a natural consequence of the three-body treatment of the hybrid meson: the larger the internal angular momenta among constituents, the more states that are generated. For these two $2^{+-}$ states, the only difference is the quantum number $L_g$, which brings similar masses but different decay modes. From Table~\ref{tab:ccgTEP}, one $2^{+-}$ state is dominated by the $D\bar{D}_2^\ast$ channel, while the other mainly decays into $D\bar{D}_1$. Since the $0^{+-}$ state here has a mass similar to the $2^{+-}$ states and also decays predominantly via $D\bar{D}_1$ channels, we recommend that experiments prioritize the $2^{+-}$ state that mainly decays into $D\bar{D}_2^\ast$. For states with ordinary quantum numbers, there are three $1^{+-}$ states, again distinguished by $L_g$. For experimental searches, we suggest the $1^{+-}$ state with mass around 4.3--4.6~GeV, as it has a significant partial decay width into the $D\bar{D}_0^\ast$ channel. For the other states with ordinary quantum numbers, once the $1^{+-}$ state is confirmed, the stable mass splittings among them may guide further measurements. Furthermore, as in our previous analysis of the $1^{--}$ charmonium hybrid with $(L_{q\bar{q}},L_{q\bar{q},g})=(1,1)$, the $c\bar{c}g$ states with ordinary quantum numbers here may also mix with nearby charmonia---an effect that can be studied within the unquenched model in future work.

\subsubsection{$c\bar{c}g$ states with $L_{q\bar{q}}=0$}

As mentioned earlier, the mainstream view in constituent gluon models is that the gauge field should carry extra angular momentum to manifest as a constituent, as exemplified by the TE-mode constituent gluon commonly adopted in current works~\cite{Qian:2026inl,Cheung:2016bym,Farina:2020slb,HadronSpectrum:2012gic,Wang:2025clb,Akbar:2024jda,Vairo:2017dmh,Lebed:2017xih,TarrusCastella:2017lex,TarrusCastella:2015mbn,Berwein:2015vca,Braaten:2014qka,Kleiv:2014kua,Sultan:2014oua,Barnes:1982tx,Chanowitz:1982qj,Barnes:1995hc,Kalashnikova:2016bta,Agaev:2025llz,Chen:2013zia,Oncala:2017hop}, especially lattice QCD~\cite{Cheung:2016bym,HadronSpectrum:2012gic,TarrusCastella:2017lex,TarrusCastella:2015mbn,Berwein:2015vca,Braaten:2014qka}. Consequently, discussions of $c\bar{c}g$ states with $L_{q\bar{q}}=0$ are rather sparse. Nevertheless, for completeness, we briefly discuss this case, in which the $L_{q\bar{q}}=0$ and $L_{q\bar{q}}=1$ configurations are considered together.

From the decay widths in Tables~\ref{tab:ccgS} and~\ref{tab:ccgP}, an important feature already emerges: nearly all these states couple strongly to ground-state charm(-strange) meson pairs. Setting aside experimental detection and viewing these results from the perspective of the unquenched quark model, we find that if the $L_{q\bar{q}}=0$ configuration is correct, mixing between these $c\bar{c}g$ hybrids with relevant charmonia could be intense, potentially leading to large mass shifts. We are currently hesitant to accept such a scenario. This issue indicates that our first-principles understanding of the constituent gluon remains incomplete, and further investigation is needed.

\subsection{A short discussion on our constituent gluon model}

Before concluding this section, we offer further remarks on our constituent gluon model. Our model is firmly grounded in the conviction that QCD is a unified theory. Thus, once we can clearly study the low-lying meson spectra, the model parameters can be directly extended to hybrid mesons. A natural question then arises: how should we understand the gluon when it manifests as a constituent? Fortunately, Refs.~\cite{Binosi:2012sj,Binosi:2022djx,Ding:2022ows} tell us that non-perturbative effects of QCD endow the gluon with an effective mass. Moreover, their computed masses for valence quarks are close to those commonly used in our potential model. Accordingly, in Ref.~\cite{Ma:2025cew}, we extended the potential model to incorporate the interaction between gluons and quarks. We found that by introducing only one new parameter, $m_g=450$~MeV, as suggested by Refs.~\cite{Binosi:2012sj,Binosi:2022djx,Ding:2022ows}, we obtained highly satisfactory results for the low-lying $1^{-+}$ light hybrid mesons. The present extension of our model to $c\bar{c}g$ hybrids with various quantum numbers has also proven successful.

However, in surveying other works~\cite{Qian:2026inl,Cheung:2016bym,Farina:2020slb,HadronSpectrum:2012gic,Wang:2025clb,Akbar:2024jda,Vairo:2017dmh,Lebed:2017xih,TarrusCastella:2017lex,TarrusCastella:2015mbn,Berwein:2015vca,Braaten:2014qka,Kleiv:2014kua,Sultan:2014oua,Barnes:1982tx,Chanowitz:1982qj,Barnes:1995hc,Kalashnikova:2016bta,Agaev:2025llz,Chen:2013zia,Oncala:2017hop,Iddir:2007dq,Iddir:2006sh}, we observe that most start from the premise that, for the lowest-mass hybrid mesons, the gluon must be in the TE mode with quantum number $1^{+-}$ to serve as a constituent, with the adopted mass typically on the order of 1~GeV~\cite{Qian:2026inl}. Since Ref.~\cite{Qian:2026inl} has shown that the TE-mode constituent gluon can be viewed as an $L_{q\bar{q},g}=1$ excitation of a $1^{--}$ gluon,we perform a naive matching: if we directly use the $1^{+-}$ constituent gluon to construct the hybrid mesons in order to reproduce the results obtained with the $1^{--}$ constituent gluon at $m_g=450$~MeV, where should the mass of this $1^{+-}$ constituent gluon lie? The answer is given in Table~\ref{tab:1+-}.

\renewcommand\tabcolsep{0.2cm}
\renewcommand{\arraystretch}{1.7}
\begin{table}[!htbp]
		\centering
\caption{Naive matching to obtain the mass of the $1^{+-}$ constituent gluon. The second column lists the new quantum numbers required to keep the original $J^{PC}$ of the $c\bar{c}g$ state; the third column gives the original mass of the $c\bar{c}g$ with a $1^{--}$ constituent gluon of $m_g=450$~MeV; the fourth column contains the reproduced masses; and the final column gives the inferred $1^{+-}$ constituent gluon mass.} \label{tab:1+-}
\begin{tabular}{ccccccccccc}
\toprule[1.pt]\toprule[1.pt]
$J^{\mathrm{PC}}$ &$(S_{q\bar{q}}, L_{q\bar{q}}, L_{q\bar{q},g}, L_{g}, L, J)$ & Origin  &Mass ($1^{+-}$) &$M_g$\\
			\midrule[1pt]
$1^{--}$  &(0, 0, 0, 1, 0, 1) & 4205  &4205 &1040\\
$0^{-+}$	&(1, 0, 0, 1, 0, 0) & 4139  &4135 &1100\\
$1^{-+}$	&(1, 0, 0, 1, 0, 1) & 4219 &4220 &1120\\
$2^{-+}$	&(1, 0, 0, 1, 0, 2) & 4241  &4248 &1050\\
        \midrule[1pt]
$0^{++}$  &(0, 1, 0, 0, 1, 0) & 4310  &4309 &990\\
$1^{++}$	&(0, 1, 0, 1, 1, 1) & 4253  &4252 &910\\
$2^{++}$	&(0, 1, 0, 2, 1, 2) & 4297  &4293 &940\\
$1^{+-}$	&(1, 1, 0, 0, 1, 1) & 4313  &4305 &980\\
$0^{+-}$	&(1, 1, 0, 1, 1, 0) & 4219  &4219 &930\\
$1^{+-}$	&(1, 1, 0, 1, 1, 1) & 4261  &4268 &960\\
$2^{+-}$	&(1, 1, 0, 1, 1, 2) & 4280  &4276 &900\\
$1^{+-}$	&(1, 1, 0, 2, 1, 1) & 4310  &4304 &1010\\
$2^{+-}$	&(1, 1, 0, 2, 1, 2) & 4315  &4315 &1000\\
$3^{+-}$	&(1, 1, 0, 2, 1, 3) & 4324  &4324 &950\\
\bottomrule[1.pt]\bottomrule[1.pt]
		\end{tabular}
\end{table}

Here, we only match the spectra with $L_{q\bar{q},g}=1$, since the $L_{q\bar{q},g}=0$ case is evidently unrelated to the TE-mode constituent gluon. Thus, as shown in Table~\ref{tab:1+-}, when the constituent gluon carries $1^{+-}$, reproducing our previous results in Table~\ref{tab:ccbarg mass spectrum} implies a mass around 1~GeV, which is precisely the value adopted in Ref.~\cite{Qian:2026inl}, albeit for a somewhat different model. In our view, this phenomenon resembles the ``overweight'' effect in classical mechanics, where a rotating object is perceived as having an enlarged mass.

As for the spin degrees of freedom of the constituent gluon, some works use a massive spin-1 gluon with three polarization degrees of freedom~\cite{Iddir:2007dq,Iddir:2006sh}, while others adopt a transverse gluon with only two physical polarizations \cite{Qian:2026inl,Farina:2020slb,Barnes:1981ac}. In this work, we use a massive spin-1 constituent gluon with three polarization degrees of freedom. However, we find that if the constituent gluon is taken to be transverse electric mode (TE mode), there will be no changes. It is because although the wave function of our constituent gluon has 3 degrees of freedom as $\boldsymbol{\epsilon}(\sigma)$, after coupling it to the $L=1$ angular momentum to be a spin-1 particle, the final wave function will approximately be
\begin{eqnarray}
    \psi(\boldsymbol{k},\lambda) \sim \sum_{m,\sigma} \langle 1, m; 1,\sigma | 1,\lambda \rangle Y_{1,m}(\boldsymbol{\hat{k}}) \boldsymbol{\epsilon}(\sigma),
\end{eqnarray}
which is just the definition of the TE mode of the constituent gluon~\cite{Qian:2026inl}, and the final expansion of it under the helicity basis through $\boldsymbol{\epsilon}(\sigma) = \sum_\lambda D^{1\ast}_{\sigma,\lambda} (\boldsymbol{\hat{k}}) \boldsymbol{\epsilon}(\lambda,\boldsymbol{\hat{k}})$ is
\begin{eqnarray}
    \psi(\boldsymbol{k},\lambda) \sim \sqrt{\frac{3}{8\pi}} \left( D_{m,-1}^{1\ast}(\boldsymbol{\hat{k}}) \boldsymbol{\epsilon}(-1,\boldsymbol{\hat{k}}) - D_{m,1}^{1\ast}(\boldsymbol{\hat{k}}) \boldsymbol{\epsilon}(1,\boldsymbol{\hat{k}}) \right).
\end{eqnarray}
Thus, nevertheless the constituent gluon has longitudinal component, after being the TE mode gluon, the longitudinal component will have no contribution to the final calculation, which means the selection rules and the number of states will keep the same.

In addition, the present three-body framework can be extended to \(b\bar b g\) hybrids by replacing \(m_c\) with \(m_b\simeq 5.10\ \mathrm{GeV}\) and refitting the ordinary bottomonium parameters. If using the current model parameters, a simple scaling estimation with the same \(m_g=0.45\ \mathrm{GeV}\) gives \(M(1^{-+})_{b\bar b g}\simeq 10.8\ \mathrm{GeV}\) and \(M(2^{-+})_{b\bar b g}\simeq 11.0\ \mathrm{GeV}\). As for their decays, similar as the charm sector, the dominant channel should be \(B\bar B_1\). However, since the our spectra are lower than the \(B\bar B_1\) threshold~(about 11~GeV), they can not decay through this channel. Thus, their final decay widths are excepted to be very small. In the present work, because we does not calculate the spectra with angular momentum $L_{c\bar{c},g}>1$, we cannot confirm the Regge trajectory behavior. A full numerical calculation is left for future work. 

\section{Summary and outlook}

As the fundamental theory of strong interactions, QCD predicts the existence of hybrid mesons with explicit gluonic degrees of freedom, which provide a crucial window for probing non-perturbative dynamics. Since the gluon can be treated as a constituent with an effective mass $m_g\approx450$ MeV~\cite{Binosi:2012sj,Binosi:2022djx,Ding:2022ows}, investigations of hybrid mesons within the framework of the potential model become possible. Furthermore, owing to the assumed unity of QCD, model parameters extracted from meson spectra calculations can be directly applied, with only one additional parameter, $m_g$, to obtain reasonable results for hybrid mesons. Our success with the $1^{-+}$ light $q\bar{q}g$ hybrid mesons indeed demonstrates the effectiveness of this approach~\cite{Ma:2025cew}.

To further validate our model, in this work we systematically investigate the mass spectra and two-body strong decay properties of ground charmonium hybrid mesons with various quantum numbers. The results show that our mass spectra are in good agreement with those of other studies~\cite{Qian:2026inl,Cheung:2016bym,Farina:2020slb,HadronSpectrum:2012gic,Wang:2025clb,Akbar:2024jda,Vairo:2017dmh,Lebed:2017xih,TarrusCastella:2017lex,TarrusCastella:2015mbn,Berwein:2015vca,Braaten:2014qka,Kleiv:2014kua,Sultan:2014oua,Barnes:1982tx,Chanowitz:1982qj,Barnes:1995hc,Kalashnikova:2016bta,Agaev:2025llz,Chen:2013zia,Oncala:2017hop,Iddir:2007dq,Iddir:2006sh}, and the predicted decay widths are also reasonable. Accordingly, based on our calculations, we recommend that experiments first search for the two $c\bar{c}g$ hybrids with exotic quantum numbers, $1^{-+}$ and $2^{+-}$, via the $D\bar{D}_1$ and $D\bar{D}_2^\ast$ channels, respectively. For the ordinary states, we suggest the $0^{-+}$ and $2^{-+}$ states, which decay predominantly into $D\bar{D}_0^\ast$ and $D_s\bar{D}_{s0}^\ast$ channels, together with the $1^{+-}$ state with a significant $D\bar{D}_0^\ast$ partial width. As for the already observed states, according to the results of both mass spectra and decay widths, \(\psi(4230)\) is the most worth tested candidate containing a hybrid component.

Although reasonable results have been obtained within our constituent gluon model, we recognize that our understanding of the constituent gluon remains incomplete. This is evidenced by the decay widths obtained for $c\bar{c}g$ hybrids in which the gauge field manifests directly as a constituent; such a configuration could have serious implications for the unquenched quark model. Therefore, for the moment, we adopt the mainstream view that the gluon must carry extra angular momentum to manifest as a constituent; in this picture, the gluon effectively acquires an enhanced mass---reminiscent of the overweight effect discussed earlier. Nevertheless, further investigation is needed, and a natural next step would be a study of $c\bar{c}$--$c\bar{c}g$ mixing.

Finally, as an outlook, the present three-body framework can be extended to \(b\bar b g\) hybrids by replacing \(m_c\) with \(m_b\simeq 5.10\ \mathrm{GeV}\) and refitting the ordinary bottomonium parameters. If using the current model parameters, a simple scaling estimation with the same \(m_g=0.45\ \mathrm{GeV}\) gives \(M(1^{-+})_{b\bar b g}\simeq 10.8\ \mathrm{GeV}\) and \(M(2^{-+})_{b\bar b g}\simeq 11.0\ \mathrm{GeV}\). As for their decays, similar as the charm sector, the dominant channel should be \(B\bar B_1\). However, since the our spectra are lower than the \(B\bar B_1\) threshold~(about 11~GeV), they can not decay through this channel. Thus, their final decay widths are excepted to be very small. In the present work, because we does not calculate the spectra with angular momentum $L_{c\bar{c},g}>1$, we cannot confirm the Regge trajectory behavior. A full numerical calculation is left for future work. 

In summary, the present work provides a unified and consistent description of the spectra and decay properties of charmonium hybrids. We hope our results will serve as a theoretical reference for future experimental searches. In particular, once exotic candidates near the predicted mass regions are observed experimentally, their decay patterns can be directly compared with our results to determine their nature as $c\bar{c}g$ hybrid mesons. This will significantly advance our understanding not only of the constituent gluon model itself, but also of the unifying and non-perturbative aspects of QCD.

\section*{Acknowledgement}
The authors want to thank Ya-Qi Cui, Ming-Hui Ding, Xiang Liu, and Ri-Qing Qian for very useful discussions. This work is supported partly by the National Natural Science Foundation of China under Grant Nos. 12305087, 12305139, 12205249, 12247101, 12475080, the Funding for School-Level Research Projects of Yancheng Institute of Technology under Grant No. xjr2022039, the General Project of Natural Science Foundation of colleges and universities of Jiangsu Province under Grant No. 24KJB140001, the Qinglan Project of Jiangsu Province, and the Xiaoxiang Scholars Programme of Hunan Normal University.

\bibliographystyle{elsarticle-num}
\bibliography{ref}

@article{ParticleDataGroup:2024cfk,
    author = "Navas, S. and others",
    collaboration = "Particle Data Group",
    title = "{Review of particle physics}",
    doi = "10.1103/PhysRevD.110.030001",
    journal = "Phys. Rev. D",
    volume = "110",
    number = "3",
    pages = "030001",
    year = "2024"
}

@article{Barnes:1981ac,
    author = "Barnes, Ted",
    title = "{A Transverse Gluonium Potential Model With Breit-fermi Hyperfine Effects}",
    reportNumber = "RL-81-017-REV, RL-81-017",
    doi = "10.1007/BF01549736",
    journal = "Z. Phys. C",
    volume = "10",
    pages = "275",
    year = "1981"
}

@article{Buisseret:2006wc,
    author = "Buisseret, Fabien and Semay, Claude",
    title = "{On two- and three-body descriptions of hybrid mesons}",
    eprint = "hep-ph/0610132",
    archivePrefix = "arXiv",
    doi = "10.1103/PhysRevD.74.114018",
    journal = "Phys. Rev. D",
    volume = "74",
    pages = "114018",
    year = "2006"
}

@article{Wang:2025dur,
    author = "Wang, Xiongfei and Liu, Xiang and Gao, Yuanning",
    title = "{Colloquium: Hadron production in open-charm meson pairs at e+e- colliders}",
    eprint = "2502.15117",
    archivePrefix = "arXiv",
    primaryClass = "hep-ex",
    doi = "10.1103/2mrp-chly",
    journal = "Rev. Mod. Phys.",
    volume = "98",
    number = "2",
    pages = "021001",
    year = "2026"
}

@article{Zhang:2025xee,
    author = "Zhang, Fu-Yuan and Huang, Qi and Wang, Li-Ming",
    title = "{Spectral analysis and decay mechanisms of 1-+ hybrid states in light meson sector}",
    eprint = "2503.01443",
    archivePrefix = "arXiv",
    primaryClass = "hep-ph",
    doi = "10.1103/1wkx-3s2l",
    journal = "Phys. Rev. D",
    volume = "113",
    number = "1",
    pages = "014002",
    year = "2026"
}

@article{Wang:2025clb,
    author = "Wang, Qian and Zhao, Qiang",
    title = "{A Short Review of the Vector Charmonium-Like State {\ensuremath{\psi}}(4230)}",
    eprint = "2508.05304",
    archivePrefix = "arXiv",
    primaryClass = "hep-ph",
    doi = "10.1088/0256-307X/42/11/110201",
    journal = "Chin. Phys. Lett.",
    volume = "42",
    number = "11",
    pages = "110201",
    year = "2025"
}

@article{BESIII:2025bez,
    author = "Ablikim, Medina and others",
    collaboration = "BESIII",
    title = "{Search for $1^{-+}$ charmonium-like hybrid via $e^{+}e^{-}\rightarrow \gamma \eta^{(\prime)} \eta_{c}$ at center-of-mass energies between 4.258 and 4.681 GeV}",
    eprint = "2504.13539",
    archivePrefix = "arXiv",
    primaryClass = "hep-ex",
    month = "4",
    year = "2025"
}

@article{Vijande:2004he,
    author = "Vijande, J. and Fernandez, F. and Valcarce, A.",
    title = "{Constituent quark model study of the meson spectra}",
    eprint = "hep-ph/0411299",
    archivePrefix = "arXiv",
    doi = "10.1088/0954-3899/31/5/017",
    journal = "J. Phys. G",
    volume = "31",
    pages = "481",
    year = "2005"
}

@article{TarrusCastella:2015mbn,
    author = "Tarr{\'u}s Castell{\`a}, Jaume",
    editor = "Andrianov, Alexander and Brambilla, Nora and Kim, Victor and Kolevatov, Sergei",
    title = "{Heavy hybrids in pNRQCD}",
    eprint = "1510.04949",
    archivePrefix = "arXiv",
    primaryClass = "hep-ph",
    reportNumber = "TUM-EFT-59-14",
    doi = "10.1063/1.4938656",
    journal = "AIP Conf. Proc.",
    volume = "1701",
    number = "1",
    pages = "050016",
    year = "2016"
}

@article{Segovia:2008zza,
    author = "Segovia, J. and Entem, D. R. and Fernandez, F.",
    title = "{Is chiral symmetry restored in the excited meson spectrum?}",
    doi = "10.1016/j.physletb.2008.02.051",
    journal = "Phys. Lett. B",
    volume = "662",
    pages = "33--36",
    year = "2008"
}

@article{Segovia:2008zz,
    author = "Segovia, J. and Yasser, A. M. and Entem, D. R. and Fernandez, F.",
    title = "{$J^{PC}=1^{--}$ hidden charm resonances}",
    doi = "10.1103/PhysRevD.78.114033",
    journal = "Phys. Rev. D",
    volume = "78",
    pages = "114033",
    year = "2008"
}

@article{Ortega:2016hde,
    author = "Ortega, Pablo G. and Segovia, Jorge and Entem, David R. and Fern{\'a}ndez, Francisco",
    title = "{Canonical description of the new LHCb resonances}",
    eprint = "1608.01325",
    archivePrefix = "arXiv",
    primaryClass = "hep-ph",
    doi = "10.1103/PhysRevD.94.114018",
    journal = "Phys. Rev. D",
    volume = "94",
    number = "11",
    pages = "114018",
    year = "2016"
}

@article{Fernandez:1993hx,
    author = "Fernandez, F. and Valcarce, A. and Straub, U. and Faessler, A.",
    title = "{The Nucleon-nucleon interaction in terms of quark degrees of freedom}",
    doi = "10.1088/0954-3899/19/12/007",
    journal = "J. Phys. G",
    volume = "19",
    pages = "2013--2026",
    year = "1993"
}

@article{Valcarce:1994nr,
    author = "Valcarce, A. and Fernandez, F. and Buchmann, A. and Faessler, A.",
    title = "{Can one simultaneously describe the deuteron properties and the nucleon-nucleon phase shifts in the quark cluster model?}",
    doi = "10.1103/PhysRevC.50.2246",
    journal = "Phys. Rev. C",
    volume = "50",
    pages = "2246--2249",
    year = "1994"
}

@article{Ortega:2016mms,
    author = "Ortega, Pablo G. and Segovia, Jorge and Entem, David R. and Fernandez, Francisco",
    title = "{Molecular components in P-wave charmed-strange mesons}",
    eprint = "1603.07000",
    archivePrefix = "arXiv",
    primaryClass = "hep-ph",
    doi = "10.1103/PhysRevD.94.074037",
    journal = "Phys. Rev. D",
    volume = "94",
    number = "7",
    pages = "074037",
    year = "2016"
}

@article{Ortega:2016pgg,
    author = "Ortega, Pablo G. and Segovia, Jorge and Entem, David R. and Fern{\'a}ndez, Francisco",
    title = "{Threshold effects in P-wave bottom-strange mesons}",
    eprint = "1612.04826",
    archivePrefix = "arXiv",
    primaryClass = "hep-ph",
    doi = "10.1103/PhysRevD.95.034010",
    journal = "Phys. Rev. D",
    volume = "95",
    number = "3",
    pages = "034010",
    year = "2017"
}

@article{Vijande:2006jf,
    author = "Vijande, J. and Valcarce, A. and Tsushima, K.",
    title = "{Dynamical study of $\bf{QQ} - \bar{u} \bar{d}$ mesons}",
    eprint = "hep-ph/0608316",
    archivePrefix = "arXiv",
    doi = "10.1103/PhysRevD.74.054018",
    journal = "Phys. Rev. D",
    volume = "74",
    pages = "054018",
    year = "2006"
}

@article{Yang:2020atz,
    author = "Yang, Gang and Ping, Jialun and Segovia, Jorge",
    title = "{Tetra- and penta-quark structures in the constituent quark model}",
    eprint = "2009.00238",
    archivePrefix = "arXiv",
    primaryClass = "hep-ph",
    doi = "10.3390/sym12111869",
    journal = "Symmetry",
    volume = "12",
    number = "11",
    pages = "1869",
    year = "2020"
}

@article{Huang:2023jec,
    author = "Huang, Hongxia and Deng, Chengrong and Liu, Xuejie and Tan, Yue and Ping, Jialun",
    title = "{Tetraquarks and Pentaquarks from Quark Model Perspective}",
    doi = "10.3390/sym15071298",
    journal = "Symmetry",
    volume = "15",
    number = "7",
    pages = "1298",
    year = "2023"
}

@article{Mathieu:2007fpv,
    author = "Mathieu, Vincent and Buisseret, Fabien",
    title = "{Short-range potentials from QCD at order $g^2$}",
    eprint = "hep-ph/0702226",
    archivePrefix = "arXiv",
    doi = "10.1088/0954-3899/35/2/025006",
    journal = "J. Phys. G",
    volume = "35",
    pages = "025006",
    year = "2008"
}

@book{Varshalovich:1988ifq,
    author = "Varshalovich, D. A. and Moskalev, A. N. and Khersonskii, V. K.",
    title = "{Quantum Theory of Angular Momentum}: {Irreducible Tensors, Spherical Harmonics, Vector Coupling Coefficients, 3nj Symbols}",
    doi = "10.1142/0270",
    isbn = "978-981-4415-49-1, 978-9971-5-0107-5",
    publisher = "World Scientific Publishing Company",
    year = "1988"
}

@article{Yang:2011rp,
    author = "Yang, Youchang and Ping, Jialun and Deng, Chengrong and Zong, Hong-Shi",
    title = "{Possible interpretation of the $Z_b(10610)$ and $Z_b(10650)$ in a chiral quark model}",
    eprint = "1105.5935",
    archivePrefix = "arXiv",
    primaryClass = "hep-ph",
    doi = "10.1088/0954-3899/39/10/105001",
    journal = "J. Phys. G",
    volume = "39",
    pages = "105001",
    year = "2012"
}

@article{Chen:2024ukv,
    author = "Chen, Xiaoyun and Tan, Yue",
    title = "{Light quarkonium and charmonium mass shifts in an unquenched quark model*}",
    eprint = "2406.00957",
    archivePrefix = "arXiv",
    primaryClass = "hep-ph",
    doi = "10.1088/1674-1137/ad53bd",
    journal = "Chin. Phys. C",
    volume = "48",
    number = "9",
    pages = "093107",
    year = "2024"
}

@article{Hiyama:2003cu,
    author = "Hiyama, E. and Kino, Y. and Kamimura, M.",
    title = "{Gaussian expansion method for few-body systems}",
    doi = "10.1016/S0146-6410(03)90015-9",
    journal = "Prog. Part. Nucl. Phys.",
    volume = "51",
    pages = "223--307",
    year = "2003"
}

@article{Hu:2020zwc,
    author = "Hu, Xiaohuang and Tan, Yue and Ping, Jialun",
    title = "{Investigation of $\Xi_{c}^0$ in a chiral quark model}",
    eprint = "2010.14984",
    archivePrefix = "arXiv",
    primaryClass = "hep-ph",
    doi = "10.1140/epjc/s10052-021-09175-9",
    journal = "Eur. Phys. J. C",
    volume = "81",
    number = "4",
    pages = "370",
    year = "2021"
}

@article{Yang:2020fou,
    author = "Yang, Gang and Ping, Jialun and Segovia, Jorge",
    title = "{$QQ\bar{s}\bar{s}$ tetraquarks in the chiral quark model}",
    eprint = "2007.05190",
    archivePrefix = "arXiv",
    primaryClass = "hep-ph",
    doi = "10.1103/PhysRevD.102.054023",
    journal = "Phys. Rev. D",
    volume = "102",
    number = "5",
    pages = "054023",
    year = "2020"
}

@article{Tan:2020cpu,
    author = "Tan, Yue and Ping, Jialun",
    title = "{X(2900) in a chiral quark model}",
    eprint = "2010.04045",
    archivePrefix = "arXiv",
    primaryClass = "hep-ph",
    doi = "10.1088/1674-1137/ac0ba4",
    journal = "Chin. Phys. C",
    volume = "45",
    number = "9",
    pages = "093104",
    year = "2021"
}

@article{Ding:2006ya,
    author = "Ding, Gui-Jun and Yan, Mu-Lin",
    title = "{A Candidate for $1^{--}$ strangeonium hybrid}",
    eprint = "hep-ph/0611319",
    archivePrefix = "arXiv",
    reportNumber = "USTC-ICTS-06-14",
    doi = "10.1016/j.physletb.2007.05.026",
    journal = "Phys. Lett. B",
    volume = "650",
    pages = "390--400",
    year = "2007"
}

@article{HadronSpectrum:2012gic,
    author = "Liu, Liuming and Moir, Graham and Peardon, Michael and Ryan, Sinead M. and Thomas, Christopher E. and Vilaseca, Pol and Dudek, Jozef J. and Edwards, Robert G. and Joo, Balint and Richards, David G.",
    collaboration = "Hadron Spectrum",
    title = "{Excited and exotic charmonium spectroscopy from lattice QCD}",
    eprint = "1204.5425",
    archivePrefix = "arXiv",
    primaryClass = "hep-ph",
    reportNumber = "TCDMATH-12-04, JLAB-THY-12-1510",
    doi = "10.1007/JHEP07(2012)126",
    journal = "JHEP",
    volume = "07",
    pages = "126",
    year = "2012"
}

@article{Qian:2026inl,
    author = "Qian, Ri-Qing and Chen, Bing and Liu, Xiang",
    title = "{Revisiting charmonium hybrid spectroscopy}",
    eprint = "2602.14766",
    archivePrefix = "arXiv",
    primaryClass = "hep-ph",
    month = "2",
    year = "2026"
}

@article{Shastry:2022mhk,
    author = "Shastry, Vanamali and Fischer, Christian S. and Giacosa, Francesco",
    title = "{The phenomenology of the exotic hybrid nonet with $\pi_1(1600)$ and $\eta_1(1855)$}",
    eprint = "2203.04327",
    archivePrefix = "arXiv",
    primaryClass = "hep-ph",
    doi = "10.1016/j.physletb.2022.137478",
    journal = "Phys. Lett. B",
    volume = "834",
    pages = "137478",
    year = "2022"
}

@article{Agaev:2025llz,
    author = "Agaev, S. S. and Azizi, K. and Sundu, H.",
    title = "{Tensor hybrid charmonia}",
    eprint = "2504.03003",
    archivePrefix = "arXiv",
    primaryClass = "hep-ph",
    doi = "10.1103/7g78-zk74",
    journal = "Phys. Rev. D",
    volume = "112",
    number = "1",
    pages = "014003",
    year = "2025"
}

@article{Kleiv:2014kua,
    author = "Kleiv, R. T. and Bulthuis, B. and Harnett, D. and Richards, T. and Chen, Wei and Ho, J. and Steele, T. G. and Zhu, Shi-Lin",
    editor = "Dasgupta, Arundhati",
    title = "{Exploring the spectrum of heavy quarkonium hybrids with QCD sum rules}",
    eprint = "1410.6259",
    archivePrefix = "arXiv",
    primaryClass = "hep-ph",
    doi = "10.1139/cjp-2014-0589",
    journal = "Can. J. Phys.",
    volume = "93",
    number = "9",
    pages = "952--955",
    year = "2015"
}

@article{Akbar:2024jda,
    author = "Akbar, Nosheen and Ali, Zeeshan and Arshad, Saadia and Irfan, Shaheen",
    title = "{Spectrum of hybrid charmonium, bottomonium, and $B_c$ mesons by power series method}",
    eprint = "2405.00350",
    archivePrefix = "arXiv",
    primaryClass = "hep-ph",
    doi = "10.1140/epjp/s13360-024-05729-4",
    journal = "Eur. Phys. J. Plus",
    volume = "139",
    number = "10",
    pages = "954",
    year = "2024"
}

@article{Sultan:2014oua,
    author = "Sultan, M. Atif and Akbar, Nosheen and Masud, Bilal and Akram, Faisal",
    title = "{Higher Hybrid Charmonia in an Extended Potential Model}",
    eprint = "1403.6941",
    archivePrefix = "arXiv",
    primaryClass = "hep-ph",
    doi = "10.1103/PhysRevD.90.054001",
    journal = "Phys. Rev. D",
    volume = "90",
    number = "5",
    pages = "054001",
    year = "2014"
}

@article{Vairo:2017dmh,
    author = "Vairo, Antonio",
    editor = "Checchia, Paolo and others",
    title = "{Quarkonium hybrids}",
    eprint = "1710.09631",
    archivePrefix = "arXiv",
    primaryClass = "hep-ph",
    reportNumber = "TUM-EFT-103-17",
    doi = "10.22323/1.314.0416",
    journal = "PoS",
    volume = "EPS-HEP2017",
    pages = "416",
    year = "2017"
}

@article{Lebed:2017xih,
    author = "Lebed, Richard F. and Swanson, Eric S.",
    title = "{Heavy-Quark Hybrid Mass Splittings: Hyperfine and ''Ultrafine''}",
    eprint = "1708.02679",
    archivePrefix = "arXiv",
    primaryClass = "hep-ph",
    doi = "10.1007/s00601-018-1376-9",
    journal = "Few Body Syst.",
    volume = "59",
    number = "4",
    pages = "53",
    year = "2018"
}

@article{TarrusCastella:2017lex,
    author = "Tarr{\'u}s Castell{\`a}, Jaume",
    editor = "Foka, Y. and Brambilla, N. and Kovalenko, V.",
    title = "{Born-Oppenheimer approximation in EFT and quarkonium hybrids}",
    doi = "10.1051/epjconf/201713706005",
    journal = "EPJ Web Conf.",
    volume = "137",
    pages = "06005",
    year = "2017"
}

@article{Berwein:2015vca,
    author = "Berwein, Matthias and Brambilla, Nora and Tarr{\'u}s Castell{\`a}, Jaume and Vairo, Antonio",
    title = "{Quarkonium Hybrids with Nonrelativistic Effective Field Theories}",
    eprint = "1510.04299",
    archivePrefix = "arXiv",
    primaryClass = "hep-ph",
    reportNumber = "TUM-EFT-45-14",
    doi = "10.1103/PhysRevD.92.114019",
    journal = "Phys. Rev. D",
    volume = "92",
    number = "11",
    pages = "114019",
    year = "2015"
}

@article{Braaten:2014qka,
    author = "Braaten, Eric and Langmack, Christian and Smith, D. Hudson",
    title = "{Born-Oppenheimer Approximation for the XYZ Mesons}",
    eprint = "1402.0438",
    archivePrefix = "arXiv",
    primaryClass = "hep-ph",
    doi = "10.1103/PhysRevD.90.014044",
    journal = "Phys. Rev. D",
    volume = "90",
    number = "1",
    pages = "014044",
    year = "2014"
}

@article{Ma:2025cew,
    author = "Ma, Zi-Xuan and Huang, Qi and Chen, Rui and Wang, Li-Ming and Tan, Yue and Hu, Xiao-Huang and He, Jun and Huang, Hong-Xia",
    title = "{Proper constituent gluon mass as the final piece to construct hybrid mesons}",
    eprint = "2504.05818",
    archivePrefix = "arXiv",
    primaryClass = "hep-ph",
    doi = "10.1103/yq7l-jmyg",
    journal = "Phys. Rev. D",
    volume = "112",
    number = "11",
    pages = "L111503",
    year = "2025"
}

@article{Farina:2020slb,
    author = "Farina, Christian and Garcia Tecocoatzi, Hugo and Giachino, Alessandro and Santopinto, Elena and Swanson, Eric S.",
    title = "{Heavy hybrid decays in a constituent gluon model}",
    eprint = "2005.10850",
    archivePrefix = "arXiv",
    primaryClass = "hep-ph",
    doi = "10.1103/PhysRevD.102.014023",
    journal = "Phys. Rev. D",
    volume = "102",
    number = "1",
    pages = "014023",
    year = "2020"
}

@article{Meyer:2015eta,
    author = "Meyer, C. A. and Swanson, E. S.",
    title = "{Hybrid Mesons}",
    eprint = "1502.07276",
    archivePrefix = "arXiv",
    primaryClass = "hep-ph",
    doi = "10.1016/j.ppnp.2015.03.001",
    journal = "Prog. Part. Nucl. Phys.",
    volume = "82",
    pages = "21--58",
    year = "2015"
}

@article{Binosi:2012sj,
    author = "Binosi, D. and Ibanez, D. and Papavassiliou, J.",
    title = "{The all-order equation of the effective gluon mass}",
    eprint = "1208.1451",
    archivePrefix = "arXiv",
    primaryClass = "hep-ph",
    doi = "10.1103/PhysRevD.86.085033",
    journal = "Phys. Rev. D",
    volume = "86",
    pages = "085033",
    year = "2012"
}

@article{Binosi:2022djx,
    author = "Binosi, Daniele",
    title = "{Emergent Hadron Mass in Strong Dynamics}",
    eprint = "2203.00942",
    archivePrefix = "arXiv",
    primaryClass = "hep-ph",
    doi = "10.1007/s00601-022-01740-6",
    journal = "Few Body Syst.",
    volume = "63",
    number = "2",
    pages = "42",
    year = "2022"
}

@article{Ding:2022ows,
    author = "Ding, Minghui and Roberts, Craig D. and Schmidt, Sebastian M.",
    title = "{Emergence of Hadron Mass and Structure}",
    eprint = "2211.07763",
    archivePrefix = "arXiv",
    primaryClass = "hep-ph",
    reportNumber = "NJU-INP 066/22",
    doi = "10.3390/particles6010004",
    journal = "Particles",
    volume = "6",
    number = "1",
    pages = "57--120",
    year = "2023"
}

@article{Bernard:2003jd,
    author = "Bernard, C. and Burch, T. and Gregory, E. B. and Toussaint, D. and DeTar, Carleton E. and Osborn, J. and Gottlieb, Steven A. and Heller, U. M. and Sugar, R.",
    title = "{Lattice calculation of 1-+ hybrid mesons with improved Kogut-Susskind fermions}",
    eprint = "hep-lat/0301024",
    archivePrefix = "arXiv",
    doi = "10.1103/PhysRevD.68.074505",
    journal = "Phys. Rev. D",
    volume = "68",
    pages = "074505",
    year = "2003"
}

@article{Dudek:2013yja,
    author = "Dudek, Jozef J. and Edwards, Robert G. and Guo, Peng and Thomas, Christopher E.",
    collaboration = "Hadron Spectrum",
    title = "{Toward the excited isoscalar meson spectrum from lattice QCD}",
    eprint = "1309.2608",
    archivePrefix = "arXiv",
    primaryClass = "hep-lat",
    reportNumber = "JLAB-THY-13-1786, TCDMATH-13-11",
    doi = "10.1103/PhysRevD.88.094505",
    journal = "Phys. Rev. D",
    volume = "88",
    number = "9",
    pages = "094505",
    year = "2013"
}

@article{Tan:2024grd,
    author = "Tan, Wei-Han and Su, Niu and Chen, Hua-Xing",
    title = "{Light single-gluon hybrid states with various exotic quantum numbers}",
    eprint = "2404.09538",
    archivePrefix = "arXiv",
    primaryClass = "hep-ph",
    doi = "10.1103/PhysRevD.110.034031",
    journal = "Phys. Rev. D",
    volume = "110",
    number = "3",
    pages = "034031",
    year = "2024"
}

@article{Barnes:1995hc,
    author = "Barnes, Ted and Close, F. E. and Swanson, E. S.",
    title = "{Hybrid and conventional mesons in the flux tube model: Numerical studies and their phenomenological implications}",
    eprint = "hep-ph/9501405",
    archivePrefix = "arXiv",
    reportNumber = "ORNL-CTP-95-02, RAL-94-106",
    doi = "10.1103/PhysRevD.52.5242",
    journal = "Phys. Rev. D",
    volume = "52",
    pages = "5242--5256",
    year = "1995"
}

@article{Chen:2013zia,
    author = "Chen, Wei and Kleiv, R. T. and Steele, T. G. and Bulthuis, B. and Harnett, D. and Ho, J. and Richards, T. and Zhu, Shi-Lin",
    title = "{Mass Spectrum of Heavy Quarkonium Hybrids}",
    eprint = "1304.4522",
    archivePrefix = "arXiv",
    primaryClass = "hep-ph",
    doi = "10.1007/JHEP09(2013)019",
    journal = "JHEP",
    volume = "09",
    pages = "019",
    year = "2013"
}

@article{Kalashnikova:2016bta,
    author = "Kalashnikova, Yu. S. and Nefediev, A. V.",
    title = "{QCD string in excited heavy-light mesons and heavy-quark hybrids}",
    eprint = "1611.10066",
    archivePrefix = "arXiv",
    primaryClass = "hep-ph",
    doi = "10.1103/PhysRevD.94.114007",
    journal = "Phys. Rev. D",
    volume = "94",
    number = "11",
    pages = "114007",
    year = "2016"
}

@article{Oncala:2017hop,
    author = "Oncala, Rub{\'e}n and Soto, Joan",
    title = "{Heavy Quarkonium Hybrids: Spectrum, Decay and Mixing}",
    eprint = "1702.03900",
    archivePrefix = "arXiv",
    primaryClass = "hep-ph",
    reportNumber = "ICCUB-17-004, NIKHF-2017-005",
    doi = "10.1103/PhysRevD.96.014004",
    journal = "Phys. Rev. D",
    volume = "96",
    number = "1",
    pages = "014004",
    year = "2017"
}

@article{Barnes:1982tx,
    author = "Barnes, Ted and Close, F. E. and de Viron, F.",
    title = "{Q anti-Q G Hermaphrodite Mesons in the MIT Bag Model}",
    reportNumber = "RL-82-088",
    doi = "10.1016/0550-3213(83)90004-4",
    journal = "Nucl. Phys. B",
    volume = "224",
    pages = "241",
    year = "1983"
}

@article{Chanowitz:1982qj,
    author = "Chanowitz, Michael S. and Sharpe, Stephen R.",
    title = "{Hybrids: Mixed States of Quarks and Gluons}",
    reportNumber = "LBL-14865",
    doi = "10.1016/0550-3213(83)90635-1",
    journal = "Nucl. Phys. B",
    volume = "222",
    pages = "211--244",
    year = "1983",
    note = "[Erratum: Nucl.Phys.B 228, 588--588 (1983)]"
}

@article{Cheung:2016bym,
    author = "Cheung, Gavin K. C. and O'Hara, Cian and Moir, Graham and Peardon, Michael and Ryan, Sin{\'e}ad M. and Thomas, Christopher E. and Tims, David",
    collaboration = "Hadron Spectrum",
    title = "{Excited and exotic charmonium, $D_s$ and $D$ meson spectra for two light quark masses from lattice QCD}",
    eprint = "1610.01073",
    archivePrefix = "arXiv",
    primaryClass = "hep-lat",
    reportNumber = "DAMTP-2016-63",
    doi = "10.1007/JHEP12(2016)089",
    journal = "JHEP",
    volume = "12",
    pages = "089",
    year = "2016"
}

@article{Iddir:2006sh,
    author = "Iddir, F. and Semlala, L.",
    title = "{Theoretical investigations on the Y(4260) being an hybrid meson}",
    eprint = "hep-ph/0611183",
    archivePrefix = "arXiv",
    month = "11",
    year = "2006"
}

@article{Iddir:2007dq,
    author = "Iddir, F. and Semlala, L.",
    title = "{Hybrid states from constituent glue model}",
    eprint = "0710.5352",
    archivePrefix = "arXiv",
    primaryClass = "hep-ph",
    doi = "10.1142/S0217751X08042687",
    journal = "Int. J. Mod. Phys. A",
    volume = "23",
    pages = "5229--5250",
    year = "2008"
}

@article{BESIII:2026mvn,
    author = "Ablikim, Medina and others",
    collaboration = "BESIII",
    title = "{Lightest $0^{-+}$ Glueball as Dominant Constituent of $X(2370)$}",
    eprint = "2607.20366",
    archivePrefix = "arXiv",
    primaryClass = "hep-ex",
    month = "7",
    year = "2026"
}

@inproceedings{Huang:2026_ICHEP_X2370,
  author    = {Yanping Huang},
  title     = {\href{https://indico.cern.ch/event/1522800/contributions/6982752/attachments/3322736/5949333/2026_ICHEP_yanping_v2.pdf}{Light Exotics at {BESIII}}},
  booktitle = {Proceedings of the 43rd International Conference on High Energy Physics (ICHEP 2026)},
  year      = {2026},
  month     = {Jul},
  address   = {Natal, Brazil}
}

\end{document}